\documentstyle[11pt,aaspp4]{article}\tighten

\lefthead{Blakeslee, Tonry, \& Metzger}
\righthead{GCs in Northern Clusters}
\def\kms{km~s$^{-1}$}

\def\hkpc{$h^{-1}\,$kpc}
\def\etal{{et~al.}} 
\def\sn{\ifmmode{S_N}\else$S_N$\fi}
\def\snr{\ifmmode{{\hbox{\sc snr}}}\else{{\sc snr}}\fi}
\def\snrz{\ifmmode{{\hbox{\sc snr}}^0}\else{{\sc snr$^0$}}\fi}
\def\vrms{\ifmmode{v^2_{rms}}\else{$v^2_{rms}$}\fi}
\def\pf{\ifmmode{{\hbox{\sc psf}}}\else{{\sc psf}}\fi}

\def\mbar{\ifmmode\overline m\else$\overline m$\fi}
\def\mibar{\ifmmode\overline m_I\else$\overline m_I$\fi}
\def\mrbar{\ifmmode\overline m_R\else$\overline m_R$\fi}
\def\mM{\ifmmode(m{-}M)\else$(m{-}M)$\fi}
\def\mo{\ifmmode m^0\else$m^0$\fi}
\def\Mo{\ifmmode M^0\else$M^0$\fi}
\def\mor{\ifmmode m^0_R\else$m^0_R$\fi}
\def\mstar{\ifmmode{m^*_1}\else{$m^*_1$}\fi}
\def\vi{\ifmmode(V{-}I)\else$(V{-}I)$\fi}
\def\viz{\ifmmode(V{-}I)_0\else$(V{-}I)_0$\fi}
\def\ri{\ifmmode(R{-}I)\else$(R{-}I)$\fi}
\def\riz{\ifmmode(R{-}I)_0\else$(R{-}I)_0$\fi}
\def\alp{$\alpha$}
\def\roneq{$r^{1/4}$}
\def\gta{\gtrsim}
\def\lta{\lesssim}
\def\phn{\phantom{1}}

\begin{document}

\title{Globular Clusters in 19 Northern Abell Clusters\altaffilmark{1}}

\author{John P. Blakeslee\altaffilmark{2},~~John L. 
Tonry\altaffilmark{3},~~{\sc and}~~Mark R. Metzger\altaffilmark{2}}
\authoremail{jpb@astro.caltech.edu}
\authoremail{jt@avidya.ifa.hawaii.edu}
\authoremail{mrm@astro.caltech.edu}
\vskip 6pt

\affil{Dept.~of Physics, 6-216, 
	Massachusetts Institute of Technology,
	Cambridge, MA 02139}

\altaffiltext{1}{Observations conducted at the
	Michigan-Dartmouth-MIT Observatory}
\altaffiltext{2}{Current address:  Palomar Observatory, California Institute
	of Technology, Mail Stop 105-24, Pasadena, CA 91125;
	jpb@astro.caltech.edu, mrm@astro.caltech.edu}
\altaffiltext{3}{Current address: Institute for Astronomy, University
	of Hawaii, 2680 Woodlawn Drive, Honolulu, HA 96822; 
	jt@avidya.ifa.hawaii.edu}
 
\begin{abstract}
We use the method developed by Blakeslee \& Tonry (1995) 
to study the globular cluster (GC) populations of
21 giant elliptical galaxies in 19~Abell clusters.
This method, applied here primarily in the $R$~band, is based
on the surface brightness fluctuations technique of extragalactic
distance measurement.
The sample galaxies range in redshift from 5000 to
10,000 \kms, and were selected primarily 
from the Lauer \& Postman (1994) survey
of brightest cluster gal\-axies (BCGs).
 
We find a tight correlation between the GC specific
frequency $S_N$ of the central bright galaxy in the cluster
and the cluster velocity dispersion.
$S_N$ also correlates well with the cluster X-ray temperature
and with the number of bright neighboring galaxies,
less well with the galaxy profile, and
only marginally with galaxy luminosity and overall cluster richness.
It does not correlate with cluster morphology class.
Thus, unlike galaxy luminosity, $S_N$
is determined by the cluster mass, or density.
To account for this situation, we propose
that the GCs formed early and in proportion to the available mass,
while the luminosity growth of the galaxy was later halted,
yielding the observed correlations of \sn\ with density.
We introduce a quantity called $\eta_{_{GC}}$, the number of
GCs per unit local cluster mass.  For a simple
cluster mass model, $\eta_{_{GC}}$ is found to be constant,
indicating a uniform GC production rate per unit available mass.

A measurement of the Gaussian width $\sigma$ of the GC luminosity
function (GCLF) is one of the byproducts of our analysis.
In the cosmic microwave background frame, the mean width 
for this sample is $\langle \sigma \rangle{\,=\,}1.43$~mag, virtually
identical to the HST value for M87, the galaxy used to calibrate the
mean of the GCLF in this analysis.

\end{abstract}

\keywords{galaxies: clusters: general ---
galaxies: distances and redshifts --- 
galaxies: elliptical and lenticular, cD ---
galaxies: star clusters ---
globular clusters: general}

\section{INTRODUCTION}

The first reported observations
of a globular cluster system (GCS), or population,
around an external galaxy was by Hubble (1932), who
``provisionally identified as globular clusters'' the relatively bright,
slightly extended objects in the halo of M31.  Progress in the 
study of extragalactic GCs was slow however, and it was more
than twenty years before Baum (1955)
identified the brightest members of the extremely rich GCS which 
surrounds M87, the central giant elliptical in the Virgo cluster.
More detailed photographic studies of this system would be another decade
in coming (Racine 1968a,b). 

The first GCS to be observed at a distance significantly beyond Virgo was
by Smith \& Weedman (1976), who reported a statistical excess of
$\sim\,$20 stellar objects around  NGC~3311, the cD galaxy at the
center of the Hydra cluster.
At $\sim\,$3500 \kms, Hydra is one of the nearest Abell clusters
(Abell 1060), and the detection by Smith \& Weedman represented 
$\lesssim\,$0.2\% of the total GC population of this galaxy.
Significant further progress in the study of GCSs, 
particularly around M87-like galaxies in clusters,
awaited advances in astronomical imaging and analysis techniques.

The revolution in astronomy sparked by CCDs reached the field
of extragalactic GC research
in the mid-1980s.  Van~den Bergh \etal\ (1985)
used the new tool to study the M87 GCS, reaching for the first time
beyond the turnover point in this galaxy's globular cluster 
luminosity function (GCLF).  CCD studies of GCs around ellipticals
in Leo (Pritchet \& van den Bergh 1985), 
Coma (Harris 1987; Thompson \& Valdes 1987), and Virgo (Cohen 1988)
followed soon after.  With the revolutionary image quality of 
the Hubble Space Telescope (HST), it is now possible to image
two magnitudes beyond the turnover in the M87 GCLF,
(Whitmore \etal\ 1995), reach unprecedented depths
along the Coma GCLF (Baum \etal\ 1995), and study a larger
number of smaller GCSs in detail (Forbes \etal\ 1996a).
 
Besides the boon to GCS research, CCDs allowed for the development
of the SBF method of distance measurement (Tonry \& Schneider 1988;
but see also Shopbell et al.\ 1993, who showed that it could be
done, with considerable effort, on photographic plates).
The SBF method measures the seeing-convolved variance, or
``fluctuations,'' produced by the
Poisson statistics of the stars in an early-type galaxy.
The amplitude of the fluctuations
decreases with the square of the distance to the galaxy,
and when divided by the galaxy's mean surface brightness, yields the
luminosity-weighted average flux of the stars within the galaxy.
This flux, usually referred to in terms of magnitudes and called
\mbar, gives the distance to the galaxy after proper calibration.
 
In the present work, the SBF image analysis methods are used to measure
the ``bumpiness'' in a galaxy image due to GCs.  Traditionally,
GC populations around galaxies have been studied through their
brightest members, which appear as an excess of faint point sources 
with roughly the same distribution as the halo light.
However, they can also be studied through the surface brightness 
variance (``bumpiness'') produced in the image by the remainder
of the population (those too faint to be detected as point sources).
This variance is a nuisance which must be subtracted from the stellar SBF
amplitude in order to derive a distance, as in the SBF survey
(Tonry et al.\ 1997). However, it can also be used as a 
probe of the GC population.
By using the counts of the brightest GCs together with the variance
from the rest of the GCs, we can constrain the luminosity function 
and determine the total GC population much more accurately,
and to much larger distances.

The number of GCs per unit {$M_V\,$}$=\,${$-$}15 of galaxy luminosity is
known as the ``specific frequency (or number) of globular clusters,''
abbreviated \sn\ (\cite{hv81}).
Some luminous central galaxies in clusters, M87 being the most famous, have
huge GC populations with \sn${\,\sim\,}$12, about three times the 
typical value for giant ellipticals.
Speculation on the origin of these ``anomalous'', or
``high-\sn'', systems has abounded,
but there have been few observational constraints.
In the first application of our analysis method (Blakeslee \& Tonry 1995;
hereafter \cite{bt95}), we studied the GC populations of the
two central giants in the Coma cluster, nearly six
times more distant than Virgo.
Due to the success of that project,
we undertook a larger, systematic study of an unbiased sample of
brightest galaxies in Abell clusters using the same technique.
Our goal was to make significant progress towards understanding the 
seemingly unpredictable variations in the observed \sn\
values of these galaxies.  Here we report on the results of that study.

Along the way, we have learned more about the GCLF, usually
represented by a Gaussian distribution in magnitudes:
\begin{equation}
N(m) \, = \, (2\pi\sigma^2)^{-\case{1}{2}}\,N_0\; e^{-(m-m^0)\over2\sigma^2} \; ,
  \label{eq:c2e3} 
\end{equation}
where $N_0$ is the total number density of GCs, $m^0$ is the apparent mean, or
``turnover'' magnitude, and $\sigma$ is the Gaussian dispersion, or ``width''.
The GCLF has been used extensively as a distance indicator based upon the
universality of the absolute magnitude of the turnover $M^0$
(see Whitmore 1996 and references therein).  
HST increases by at least a factor of five the maximum distance to which
this method can be used and may soon resolve the lingering questions
regarding possible variations in $M^0$.
In our initial Coma study, we extended by a factor of five the distance 
to which GCLF widths had been measured,
and subsequent work reported here has gone further still.

In the following section we discuss the sample selection for this
study and describe the observations.  In Section~3, we detail the
data reduction and analysis procedure, essentially the same as that
of \cite{bt95}.  Section~4 presents the results on the specific
frequencies and GCLFs of the sample galaxies, and Section~5 explores
the relationship of \sn\ to other properties of the galaxies and the
galaxy clusters.  We find the first clear evidence for strong
correlations of \sn\ with cluster properties; some of which
were presented previously by Blakeslee (1997a).  In Section~6,
we compare our results to those predicted by various models and 
attempt to synthesize the simplest scenario consistent with 
observations.  The final section summarizes this work.

\vfill\eject

\section{SAMPLE SELECTION AND OBSERVATIONS}

\subsection{Abell Cluster Galaxy Sample}\label{sec:sample}

The galaxy sample used in this study was selected from the Lauer \&
Postman~(1994; hereafter LP) volume-limited survey of
brightest cluster galaxies (BCGs) in 119 Abell and ACO galaxy clusters
(Abell 1958; Abell, Corwin, \& Olowin 1989).  The LP sample includes
nearly all the Abell/ACO clusters with redshift $cz<15,000$ \kms\ and 
galactic latitude $|b| > 15^{\circ}$.  Seven clusters were excluded 
from their survey because of spiral, star-forming irregular, 
or anomalously faint BCGs,
six ``clusters''  were not true galaxy over-densities,
and  two were not observed by LP.
Postman \& Lauer (1995; hereafter PL) present and discuss the data used
by LP, including observations of second brightest, or second
ranked, galaxies in clusters for which the magnitude offset
between first and second was small.

To ensure that reasonably accurate statistical statements about GCSs
would result, we selected as complete a subsample as possible
from the LP survey.  The final sample studied
here includes all of the northern hemisphere 
LP and PL galaxies with $cz < 10,000$ \kms\ and
$\alpha>0.4$, where $\alpha \equiv d\,$log$\,L/d\,$log$\,r|_{r_m}$
is the logarithmic slope of the galaxy luminosity evaluated at the metric
radius $r_m{\,=\,}10$~\hkpc.  (Larger \alp\ implies a more extended galaxy;
it was first introduced by Hoessel [1980], who called it the
``structure parameter'' of the galaxy profile).
{In~addition,} NGC~4839 in Coma has been added to this sample because
it is a known cD galaxy, though not the first or second brightest Coma member.
(A cD galaxy is a giant elliptical that possesses an extended, low surface
brightness halo; its radial profile exhibits a significant excess
of light relative to an \roneq\ law. See Tonry [1987] for a review.)
Thus, the sample comprises 23 galaxies in 19 Abell clusters,
two of which are the Coma galaxies studied by \cite{bt95}.
Of the nine true clusters omitted by LP, only A426
(the Perseus cluster, with its giant disturbed BCG Perseus~A)
and A400 (which LP reject as ``anomalously faint'') 
are within the above redshift and coordinate cutoffs.

Table~\ref{tab:bcg} lists the galaxies in the present sample.
The columns are: galaxy identification (Abell number of, and galaxy rank
in, cluster); right ascension and declination (J2000 coordinates);
galactic longitude and latitude;
heliocentric \hbox{velocity} in \kms; $B$-band extinction (as listed by
Burstein \& Heiles 1984); $R$-band absolute metric magnitude and
\alp\ parameter (both as
listed in LP and PL, except A1656-3, 
for which the values were determined from the data presented here);
and the ``common name'' for each galaxy.  The common galaxy name was
chosen first to be the NGC or IC number (prefixed by N or I, respectively),
then the UGC number (Nilson 1973; prefixed by U), and last by the CGCG
number (Zwicky 1968; prefixed by C).
The table also notes the galaxies found by Schombert (1988) 
to possess cD envelopes.

Table~\ref{tab:cls} contains reference
information on the galaxy clusters themselves.  For each cluster, the 
table lists J2000 coordinates of the centroid of the extended
X-ray emission (Jones \& Forman 1997); mean
velocities in the heliocentric, cosmic microwave background (CMB), and
``Abell cluster inertial'' (ACI) frames (\kms; from PL and Postman 1996);
velocity dispersion of the member galaxies
(\kms; described in \S \ref{ssec:corr}); X-ray gas temperature
(keV; from Jones \& Forman 1997); richness class and Bautz-Morgan type (Abell 
\etal\ 1989); and Rood-Sastry type (Struble \& Rood 1987).
The ACI frame is the frame in which LP determined their clusters to have
no net peculiar velocity according to the $L_m$-\alp\ distance indicator.
The only cluster in the sample with a colloquial name is Abell~1656,
the Coma cluster.

\subsection{Observing Procedure and Runs}

The data used in this study were obtained with the 2.4~m telescope at the
Michigan-Dartmouth-MIT (MDM) Observatory on Kitt Peak over the course
of several observing runs.  
The integration times for the individual exposures on the program galaxies
ranged from 10 to 15 minutes.
Between exposures, the telescope was shifted by 5--10$^{\prime\prime}$ 
in order to improve the image flattening, allow for the removal of cosmic
ray hits and CCD defects, and provide a check on whether a series of
exposures was truly photometric.  The total integration times on the
program galaxies ranged from just over 1~hr to nearly 6.5~hr,
determined mainly by the galaxy's distance.  
Each night of observing, approximately 10 twilight flat fields
were taken through the filter which was to be used for the deep imaging
(either $R$ or $I$). If the sky was clear,
about 10~images of Landolt (1992) photometric standard star fields
were taken at varying airmasses throughout the night
in each band for which photometry was needed.
Standard star reductions and  photometric calibration of the final galaxy
images are discussed in  \S\ref{ssec:phot}.

Table~\ref{tab:runs} summarizes the information on all the observing
runs. The runs are designated by the month and year in which they took place.
For each run, the table lists
the CCD camera that was used, the pixel scale of the images in
arcseconds per pixel, the filter through which the GC data were taken, 
the typical photometric coefficients ($m_1$, $A$, and $C$, defined below
in Eq.~[\ref{eq:phot}]) 
used in calibrating the GC data, and some notes.
(The CCD listed as ``STIS 2048$^2$'' was fabricated as part of the
STIS program and obtained by MDM, although the STIS chip later
installed on HST is a 1024$^2$ device.)
Collecting data for this project was not the primary goal of either of
the first two runs. 
Only the two Coma galaxies from \cite{bt95} were observed
in the 0593 run, and only one sample galaxy was observed during 0794.
The 0593 images were binned 2$\times$2 to yield 0\farcs343~pix$^{-1}$.
After imaging in the $I$~band for the first two runs,
we chose to take the remainder of the data in $R$ for several reasons:
(1)~The $R$ band is near the peak sensitivity of the CCD,
(2) thinned chips like the MDM Tek 1024$^2$ (the detector of choice for its
high quantum efficiency) show significant fringing in~$I$,
(3) the stellar SBF is relatively stronger in $I$, with
\mrbar$-$\mibar${\,\sim\,}1.6$, while $(R{-}I)\sim0.5$ for GCs,
(4) the sky is much brighter in $I$ than $R$, and
(5) PL report precise aperture photometry in $R$, which can be used
as a cross-check for our photometry (Blakeslee \etal\ 1998).
A glass $R$ filter provides a much better match to the standard bandpass,
so it was used in lieu of an interference filter; the reverse is true
in the $I$~band, where the open red end of glass filters makes them
a poor choice for CCD photometry.  
 
\section{REDUCTIONS AND ANALYSIS}

\subsection{Image Processing and Calibration}\label{ssec:phot}

High signal-to-noise flat fields were produced for each night as follows.
The individual frames were bias corrected by fitting the overscan
region of the CCD (which had variable
top-to-bottom structure in the bias) to a $\sim\,$5th order spline
and subtracting.   Each frame was then windowed and
divided by the straight sum of all the frames, allowing any stars 
present to be easily identified and masked out.
The sum was then multiplied back in and deviant pixels due to
cosmic rays hits were identified by an automatic procedure 
(Tonry \etal\ 1997).  The pixels affected by the cosmic rays and
stars were then replaced by scaled data values determined according
to the pixels' values in the rest of the image stack.
The frames were then summed to to produce the final flat field.
The galaxy data and standard star calibration frames were bias subtracted
and windowed in the same manner as the flats and divided by the
normalized flat field.

The Landolt standard star fields used in this project are mostly
the same as those listed by Tonry \etal\ (1997).
They were selected because of the large number of observations 
made by Landolt on each, and consequent small magnitude uncertainties, and 
because of the number of stars present within a
$\sim5^\prime{\times}5^\prime$ field.
Following Landolt (1992), and as in Tonry \etal\ (1997), 
we summed the flux from each photometric standard within an aperture
of 7\arcsec\ radius.  Standard star measurements
with estimated errors larger than 0.02~mag from flux, sky,
and Landolt magnitude uncertainties were discarded.  After
all the standard star observations from a run were reduced, 
the results were fitted according to 
\begin{equation}
	m = -2.5\log(f/t) -A\, \sec(z) + C\, \vi + m_1
	\label{eq:phot}
\end{equation}
where $m$ is the magnitude reported by Landolt in the appropriate band,
$f$ the total counts in electrons from the star within 7\arcsec,
$t$ the exposure time in seconds, $z$ the zenith angle, and \vi\
comes from Landolt.  Typical values of $A$, $C$, and $m_1$ were
given in Table~\ref{tab:runs} for each run.
On truly clear nights, the scatter about the derived relation
was $\lta\,$0.01~mag.

The individual exposures making up
a galaxy observation were bias subtracted and flattened before being
brought into registration based on the positions of stars in the images.
Bad pixels were masked, and the frames were then combined, rejecting 
cosmic ray hits, to make the final galaxy image.
The following two subsections discuss the properties of the 
images which make up the present data set.

\subsection{Image ``Quality''}\label{ssec:qual}

One measure of the depth of an observation is the quantity \mstar, the
magnitude of an object which will produce one count per total image
integration time, corrected for Galactic extinction:
\begin{equation}
	\mstar = 2.5\log(t) -A\, \sec(z) + C\, \vi + m_1 - A_\lambda \, ,
	\label{eq:mstar}
\end{equation}
where $A_\lambda$ is the extinction in the $\lambda$ band.
Thus, the extinction corrected magnitude of an 
object yielding $f$ total counts in the image can be calculated as:~ 
$m = -2.5\log(f) + \mstar$.  So, \mstar\ indicates the amount of
signal in an image, but takes no account of the noise.

It would be helpful to have some measure of the quality of an image,
i.e., the suitability of an observation for
yielding the desired information.  
The approximate signal-to-noise ratio (\snr) of a sky-limited object
producing $f$ counts within the \pf\ full width at half maximum (FWHM) is
\begin{equation}
	\snr = {f \over {\sqrt{b \times \pf^2}}}  \, ,
	\label{eq:snr1}
\end{equation}
where $b$ is the sky level in counts/arcsec$^2$ and \pf\ is the
FWHM in arcsec.  (To calculate the true signal-to-noise, the galaxy
background would also need to be taken into account.)
For a given luminosity, such as a fixed point along
the GCLF, the flux received by the detector scales
as $f \sim t/d^2$, where $d$ is the distance to the source.
Thus, in order to reach the same \snr\ at this fixed luminosity,
the exposure time must scale as $t \sim d^4$.
Exposure times for our sample galaxies were scaled in this way
(taking into account sky background and seeing variations), with
the goal being to reach within $\sim$2~mag or so of the expected
GCLF turnover \mo\ while the point source identification was
still $\gta\,$90\% complete.

It is interesting to view Eq.~(\ref{eq:snr1}) in terms of magnitudes:
\begin{equation}
2.5\log(\snr) = -2.5\log(\pf) - m_{\rm FW} + 0.5\,(\mstar + \mu_{sky})
 \, ,
\label{eq:simple}
\end{equation}
where $m_{\rm FW}$ is the magnitude of the object within the 
\pf\ FWHM and $\mu_{sky}$ is the sky surface brightness in mag/arcsec$^2$.
For a \pf\ which approximates a two dimensional Gaussian, the total
flux is about twice the flux within the FWHM.
The obvious characteristic magnitude for GC studies
is \mo, the point at which the GCLF turns over.
The limiting magnitudes of the images in this project do not 
generally approach \mo;  however, if we plug in $m_{\rm FW}{=}m^0$,
the object under consideration will be one which is twice as
bright as \mo, e.g., a globular cluster 0.75~mag brighter than the GCLF
turnover.  Eq.~(\ref{eq:simple}) then becomes:
\begin{equation}
2.5\log(\snrz) = -2.5\log(\pf)  + 0.5\,(\mstar + \mu_{sky}) - \mo \, ,
\label{eq:qual}
\end{equation}
where we have called this fiducial signal-to-noise ratio \snrz, although
it actually refers to an object twice as bright as, not equal to, \mo.
In 1\arcsec\ seeing, the quantity calculated in Eq.~(\ref{eq:qual}) 
is positive if the average of \mstar\ and $\mu_{sky}$ is fainter
than the GCLF turnover \mo\ in the galaxy.  Perhaps
more intuitive, however, 
is \snrz\ itself:
\begin{equation}
\snrz = \pf^{-1} \times 10^{0.2\,( \mstar + \mu_{sky} - 2\,\mo )} \, .
\label{eq:snrz}
\end{equation}
This is a useful indicator of the data image quality once
\mo\ has been estimated.   

\def\dmcv{$\Delta{(m{-}M)}_{CV}$}

We can estimate \mo\ from its measured value in the Virgo cluster, 
first placing Virgo into the CMB frame using the relative distance
between it and one of the sample clusters.  From the review by van den Bergh
(1992), we adopt \dmcv $\,= 3.71{\,\pm\,}0.10$ mag as the relative 
Coma-Virgo distance modulus.
Recently, Baum \etal\ (1997) have found \dmcv$\,= 4.10{\,\pm\,}0.06$ mag
from a maximum likelihood comparison of HST observations of the GCLF in
IC~4051 with the HST GCLF from Whitmore \etal\ (1995).  They also cite the
work of Giovanelli (1996) as indicating \dmcv $\,= 3.95$ mag.
However, the large, homogeneous ``Mark~III'' Catalog (Willick \etal\ 1997)
of Tully-Fisher distances yields \dmcv $\,= 3.66{\,\pm\,}0.15$ mag
(Willick 1997), and a re-investigation 
of the $D_n$--$\sigma$ and fundamental plane relations in the two clusters
finds \dmcv $\,= 3.55{\,\pm\,}0.15$ mag (D'Onofrio \etal\ 1997).
The value we have adopted 
is in agreement with these determinations, while values
near 4.0~mag are not.  Thus, using the CMB velocity of Coma
in Table~\ref{tab:cls}, we have:
\begin{equation}
\mo(z) = \mo_{\rm Vir} + 5\,\log\left(cz\over1310\right).
\label{eq:m0ofz}
\end{equation}
This calibration is discussed further in \S\ref{ssec:constrain}.

\subsection{Final Data Set}\label{ssec:data}

Table~\ref{tab:obs} presents the data set used for this study of the
GCSs of Abell cluster galaxies.  For each galaxy image, the table lists
the observing run in which the data were taken,
the total exposure time in seconds,
the seeing FWHM in arcsec, \mstar\ as defined in Eq.~(\ref{eq:mstar}),
the sky brightness in mag/arcsec$^2$, \snrz\ calculated as described
above, and the total number of unique objects identified in the image 
down to a signal-to-noise limit of 4.0 by the photometry program DoPhot
(Schechter, Mateo, \& Saha 1993; DoPhot and the point source photometry 
will be discussed in \S\ref{ssec:doph}).   
Galaxies A347-1 and A569-1 were both observed twice, and each observation
was reduced independently.  For these two galaxies the strong dependence
of the number of identified objects on \snrz\ is clearly evident.

Of course, \snrz\ does not tell the whole story.  The total signal-to-noise
of the data
scales also with the square~root of the number of GCs present in the
image.  This cannot be known ahead of time, though some estimates
might be made based on galaxy luminosity.   For comparison purposes,
Table~\ref{tab:obs} includes the information on the Coma observations
presented and analyzed by \cite{bt95}.  Both of these galaxies
happened to have very rich GC populations. Consequently, it was possible
to learn a good deal about them, despite the fact that their \snrz\ values 
are among the lowest in the table. (Although the comparison is not really
fair, as those images were in the $I$-band where the sky is much brighter
but the galaxy background is not, and we have neglected the galaxy
background here.)

\subsection{Point Source Reductions}\label{ssec:doph}

\subsubsection{Producing a Smooth Background}

\begin{figure}\epsscale{0.85}\plotone{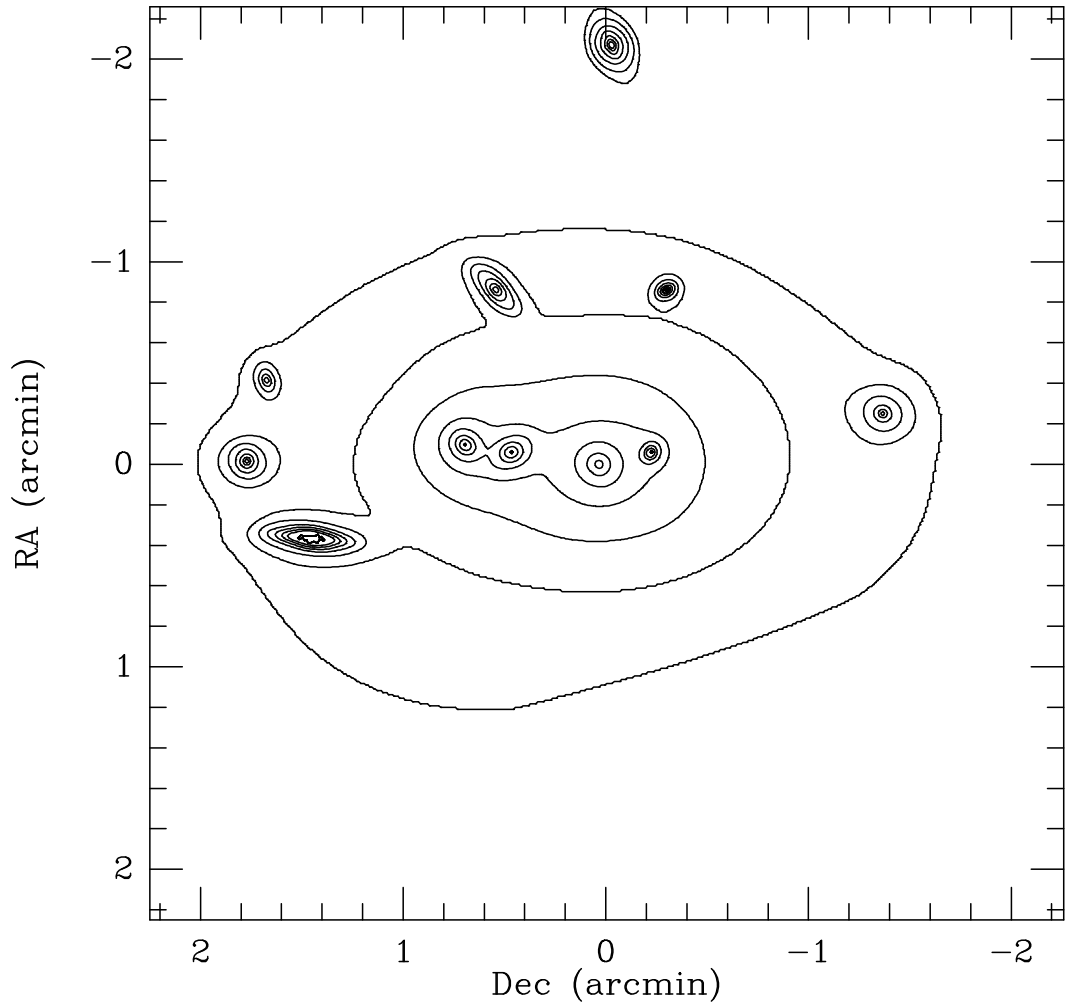}
\caption{
The isophotal model for the complex A539-2 galaxy system,
generated by the iterative fitting procedure described in the text.
The isophotal contours are plotted in increments of 1~mag, with the outermost
contour at $R = 24.5$ mag/arcsec$^2$.
\label{fig:a539-2.mod}}
\end{figure}

As in \cite{bt95}, we mask out all the easily visible stars
and small galaxies before fitting a smooth model to the galaxy light.
For many of these fields, the main galaxy and several smaller
galaxies were modeled and subtracted using an iterative procedure.
In some of the images, however, the multiple galaxies are comparable in size,
close together, and show interaction.    
For these galaxies, a single smooth galaxy model could not be 
fitted at the first iteration; thus, we used the simultaneous nucleus
fitting software developed by Lauer (1988) and also used by LP. 
After simultaneously modeling
and subtracting the primary galaxy and its secondary nuclei, we added the
primary back in and could then successfully model it with our software.
The same was done for the rest of the nuclei until all of them had
been modeled and subtracted, minimizing the model residuals.
An extreme case can be seen in Figure~\ref{fig:a539-2.mod}, which 
shows the model generated in this way for A539-2.

After subtracting the final galaxy model, we mask all visible objects
and fit the large scale model residuals to a grid of spacing roughly
10~times the size of the \pf, interpolating between grid points.
Subtraction of the large scale residuals leaves a
very flat ``residual image'' on which the point source photometry
and power spectrum measurements described below are carried out.

\subsubsection{Finding the Objects}

The automatic photometry program
DoPhot (Schech\-ter \etal\ 1993) was run on an integer version of each
residual image in which saturated stars, ``large'' dwarf galaxies,
and other bad features had been masked. 
Afterwards, DoPhot's model image (generated by taking the difference 
of the input and output images) was closely inspected
for extended objects which had been fitted as tight bunches
or strings of point sources.  Objects larger than DoPhot's fit box cannot
be fitted as single objects, but even some smaller ones refuse to conform
to any reasonable model DoPhot might try.  Thus, in all cases, it was
necessary to mask out these problematic objects (which sometimes included
residuals from the galaxy subtraction) and re-run DoPhot.
Aside from these aberrant cases,
DoPhot does a good job of distinguishing between point sources
and extended objects, as tests performed by Ajhar, Blakeslee, \& Tonry
(1994) showed.  This proves to be useful for rejecting
possible dwarf galaxies, perhaps as faint as the brightest GCs,
which may cluster around the BCG.

DoPhot's fit magnitudes must be calibrated onto an absolute scale. 
This is done by taking the median difference of the aperture and
fit magnitudes for the $\lta\,$20 brightest objects.  
Judging from the scatter among these objects, the uncertainty in
this correction is 0.02--0.03~mag, or about twice as large as the
photometric uncertainty from the standard star calibration.
The total number of unique objects identified and fitted by DoPhot
in each field were listed in Table~\ref{tab:obs}.
These numbers do not precisely reflect the relative
densities of objects among the images because the final field size
varies due to unequal telescope shifting, windowing, and masking 
(as well as different detectors in a few cases), but they convey a
general impression of the number of objects present in each residual image.  

\subsubsection{Completeness}\label{ssec:comptests}

In order to determine the completeness of DoPhot in finding point
sources as a function of magnitude, a bright, but not saturated,
isolated star was chosen
from each field and cloned into a grid of stars with separation
$\sim\,$10\arcsec.  
The grid was then scaled 5--7 times in 0.5~mag steps, with
the noise in the grid increased appropriately for each scaling.
The brightest scaling was at $m{\,=\,}$22.0--23.0, depending on the
depth of the image.
Each scaled grid of stars was added to a separate copy of the real
data, and DoPhot was run on the image; all parameters were set
to the same values as in the original run. Then the results of each run
were matched to the grid stars to determine the completeness
and any possible magnitude bias.

The test results for each field were used in choosing the
cutoff magnitude $m_c$ at which the point source completeness
was $\sim\,$90\% and the photometric error small.  Actually, the completeness
depends on radius, so typically two different cutoff magnitudes
were used, with $m_c$ being 0.5~mag fainter outside the central
${\sim\,}1^{\prime}$.  For each separate region of the image,
an uncertainty in the completeness fraction 
$f_c \equiv {\rm N}_{\rm found}/{\rm N}_{\rm add}$ was calculated as
$\delta f_c = [f_c\,(1{\,-\,}f_c)/{\rm N}_{\rm add}]^{1/2}$ (Bolte 1989).
In general, the corrections to point source counts are small 
($<10\%$); more detail on the completeness tests is given 
by Blakeslee (1997b).

\subsubsection{Background Counts}

After rejecting extended objects  and
correcting for incompleteness, we fitted the radial distribution
of point sources in each field to an $r^{1/4}$~law plus background model.
The primary reason
for this procedure was to determine the background contamination from
unresolved galaxies or faint stars.   Many different
binnings were explored for each field, and an average
background value was chosen from among the fits.
When the innermost points significantly changed the fit (due to
a leveling off of the counts at small radii), they were excluded
(similar to the power spectrum fits in \S\ref{ssec:powspec}).
A supplementary paper (Blakeslee \etal\ 1998) will discuss in more
detail the point source photometry and radial distributions, including
a comparison to the halo light distributions,
along with similar data from a sample of southern BCGs.
In the following section we discuss the power spectrum measurements
and background estimates, and in \S\ref{ssec:constrain} we tabulate
the point source densities and power spectrum normalizations, 
including background corrections.

\subsection{Fluctuation Reductions}

\subsubsection{Measuring the Power Spectra}\label{ssec:powspec}

We use the same power spectrum analysis method as was used by \cite{bt95},
and described in more detail by Tonry \etal\ (1990).
After all objects brighter than the cutoff magnitude $m_c$ are masked out, 
the image power spectrum $P(k)$ is modeled as a linear 
function of the ``expectation power spectrum'' $E(k)$:
\begin{equation}
\label{eq:power}
P(k) = P_0 \times E(k) + P_1.
\end{equation}
$E(k)$ is computed as the convolution of the power
spectra of the \pf\ and the window function of the mask.
(The mask gets multiplied in image space, convolved in Fourier space.) 
$P_0$ is called the fluctuation power, the spatial variance 
in intensity which has been convolved
with the \pf\ and therefore must have originated above the atmosphere.
$P_1$ is the white noise component, which must be overcome by
the signal.
The star used for modeling the power spectrum of the \pf\ in each image
was the same one used for doing the completeness tests.

The power spectra frequently show excess power at low wavenumber
due to shells, tidal distortions, etc.~which were not removed by the
galaxy modeling, as well as from imperfect
flattening at very low wavenumber.
To deal with this problem, the low wavenumbers were omitted from the
fits used to determine $P_0$.  
In order to determine the lowest usable (i.e., uncontaminated) 
wavenumber $k_L$, we fitted Eq.~(\ref{eq:power}) to the data power
spectrum over the interval $(k_{min}, k_{max})$, setting $k_{max}$ 
to the maximum wavenumber in the data and varying $k_{min}$.
The point at which the $P_0$ vs.\ $k_{min}$ relation
flattened out was adopted as $k_L$.
Then, in an effort to eliminate some of the arbitrariness in the choice
of $k_L$, we took a weighted average of the derived
$P_0$ values from among the fits that had lowest wavenumbers ranging
from $k_{min} = k_L$ to $k_{min} \approx {\,2\,}k_L$;
the rms variance in  $P_0$ over these fits was used as an estimate of
the uncertainty due to the choice of $k_L$. 
$P_1$ is so well determined by the high wavenumbers that it
changes only negligibly among these fits.

\begin{figure}\epsscale{0.85}\plotone{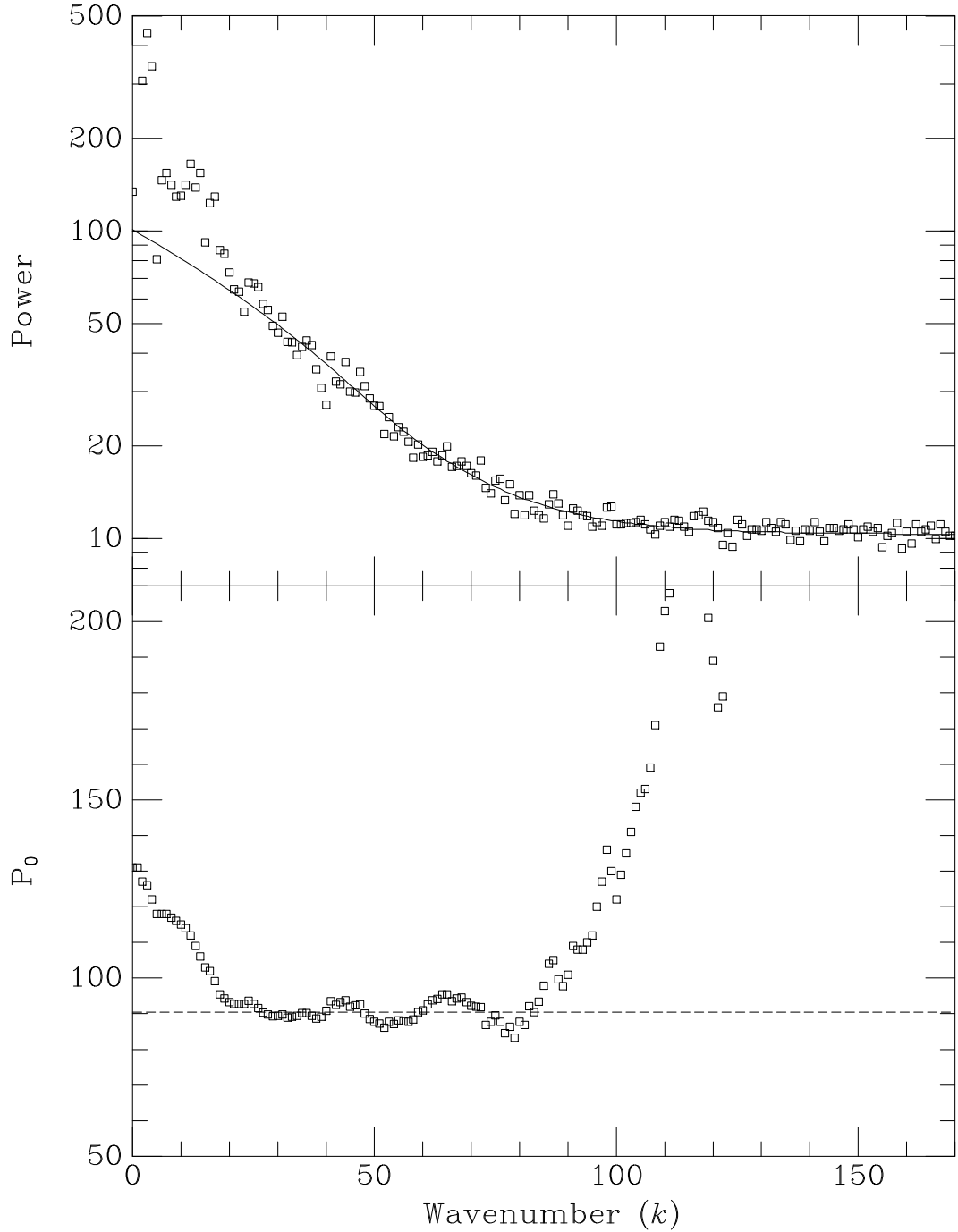}
\caption[fig02.ps]{
Power spectrum of the A2197-2 residual image for an annulus 
extending from 128 to 256 pixels in radius $(a)$, and
the power spectrum normalization $P_0$ as a function of the starting
wavenumber of the fit $(b)$.
The solid curve is an example fit to the power spectrum, and the 
dashed line is the final value chosen for $P_0$.
\label{fig:a2197-2pow}}
\end{figure}

Figure~\ref{fig:a2197-2pow} provides an example of the power spectrum fits.
The upper panel shows the power spectrum of an annulus of the A2197-2 image;
the lower panel shows the fitted $P_0$ as a function of the lowest 
wavenumber used in the fit. (The power has been divided by the mean galaxy 
intensity to yield more manageable numbers.)  
The fits used in determining $P_0$ had lowest wavenumbers ranging
from 32 to 65.
The dashed line in the lower panel indicates the final value $P_0$
derived from an average of these fits.

\subsubsection{Globular Cluster and Background Fluctuations}

The mathematical formalism for this section was given by \cite{bt95}; here,
we describe the corrections for background variance applied
in this sample.  Two sources of contamination must taken into account:
the stellar SBF and the fluctuations due to background galaxies.
These must be subtracted from the $P_0$ measurements
described in the previous section in order to determine the amplitude
of the fluctuations due to GCs.  

Most of the observations reported here are in the $R$ band, unlike the
Coma $I$-band observations of \cite{bt95},
but the correction for the stellar SBF is done in
the same way, relative to the measured value of the fluctuation magnitude
in Virgo, where \mrbar $=$31.25 for the BCG NGC~4472 and other big
ellipticals (Tonry \etal\ 1990).   We include a generous allowance of
0.30~mag for the uncertainty in \mrbar\ (intrinsic variations in
\mrbar\ have not been well studied).   The contribution from the
uncertainty in the relative distances with respect to Virgo is
included in the discussions of the calibrational errors 
in \S\ref{ssec:widths} and \S\ref{ssec:sn}.
  
While the mean \ri\ color of the GCs is about 0.55
(Ajhar, Blakeslee, \& Tonry 1994), the galaxy color 
is \ri $\,\approx\,$0.68, and the color of the SBF is
(\mrbar{$-$}\mibar){$\,\approx\,$}1.6 (Tonry \etal\ 1990).
Thus, for a fixed value of 
$(m_c{\,-\,}m^0)$, the ratio of the power produced by
faint GCs to the power from the stellar SBF improves by a factor of
\begin{eqnarray}
	\left(P_{GC}\over P_{SBF}\right)_R
	\times \left(P_{GC}\over P_{SBF}\right)_I^{-1}
	& = & 10^{\,-0.4\,[\,2\,\ri_{\rm GC} 
	- (\overline m_R{-}\overline m_I) - \ri_{\rm gal}\,]} \nonumber \\
	&\approx & 3.0\, .
	\label{eq:gcfimprove} 
\end{eqnarray}
(See Eqs.~[10] and [12] of \cite{bt95}.)
The GC fluctuations in the $R$~band are roughly three times stronger
relative to the SBF than they are in the $I$ band.   
Otherwise, this ratio of fluctuation powers scales with $m_c$ and
distance modulus as shown in Figure~1 of \cite{bt95}.

We included $K$-corrections in our estimates of the
fluctuation magnitudes to account for the slightly different 
region of the rest frame spectrum that passes through the imaging filter
as a function of galaxy redshift.
The corrections were based on calculations done by Worthey (1996),
who kindly redshifted his published stellar populations models (Worthey 1994).
Unlike for the SBF Survey (Tonry \etal\ 1997), where 
$K_I(z){\,=\,}7{\,\times\,}z\,$ was used for the \mibar\ $K$-correction,
the correction here was not well described by a linear function 
out to the redshift limit of this sample; however, a quadratic form
proved adequate.  The effect of the $K$-correction in the $R$~band
is to make the estimate of the contamination
from the stellar SBF smaller by 8\% at 5000~\kms\
and by 13\% at 10,000 \kms, the limit of this survey.
In practice, the stellar SBF contamination is only $\sim\,$10\%,
so this small correction to it does not change the final results in any
significant way, but it is important for avoiding bias.
The uncertainty in the $K$-correction based on the scatter in 
Worthey's models was included in the calculation of our errors.

The larger correction 
($\sim\,$10--50\%, depending on $(m_c{-}m^0)$ and radius in the galaxy)
to the measured variance
is the one applied to account for the faint background galaxies.
As in the SBF Survey and \cite{bt95}, this is done 
by extrapolating the results from
maximum likelihood routine (to be described 
by Ajhar \etal\ [1997])  which fits
the magnitude distribution of the galaxies to a power law
in the outer parts of the image, using the observed radial variation
in the counts to subtract off the GCs.
This background estimation can be a difficult and uncertain procedure when
the galaxy nearly fills the image, but in such cases the relative size
of the correction will be smaller.
Out of necessity, the slope of the galaxy
magnitude distribution is taken from Tyson (1988), and the normalization 
$T_n$ is fitted relative to his counts over the faintest few magnitudes
of the image, in order to avoid possible dwarf satellites of the central 
galaxy.  (Example outputs from the maximum likelihood program  were shown by
by Blakeslee [1997b]).  From these measurements, the variance due to
background galaxies is calculated according to Eq.~(9) of \cite{bt95}
and subtracted along with the SBF estimate from the measured variance $P_0$
in order to obtain $P_{GC}$, the variance due to GCs. 

There is one other source of contamination to the variance measurement
which one might consider, namely the possibility that a concentration of
very faint dwarf galaxies may surround the BCG and contribute to $P_0$.
For such dwarfs to cause problems, they would have to have
a spatial distribution similar to the GCs (so as not to mimic the
background galaxies) and a distribution in
apparent magnitude extending fainter than $m_c$.
For our sample, $(m_c{-}m^0)$ is typically $\sim\,$2.5~mag;
thus, the dwarfs would have to
be fainter than about $M_R{\,=\,}{-}10.5$ (adopting a distance scale as
in \S\ref{ssec:sn}).
The only known galaxies of such low luminosity are 
diffuse, low surface brightness dwarf speroidal (dSph)
members of the Local Group (and possibly
one faint member of Virgo, outside the cluster core [Durrell 1997]);
these objects would surely be destroyed in the halo of a giant galaxy
at the center of a rich cluster.

Coma is the only Abell cluster with a well studied faint-end
of the galaxy luminosity function, and the data are basically similar to
what is observed in Virgo (e.g., Ferguson \& Binggeli 1994). 
Thomas \& Gregory (1993) and Lobo \etal\ (1997) studied the dwarf
galaxy luminosity function down to about $M_R{\,=\,}{-}16$ 
and $M_R{\,=\,}{-}14$,
respectively.  They both concluded that down to these luminosities,
the dwarf elliptical (dE) and dSph populations followed the same spatial
distribution as the larger cluster galaxies, except in the cluster core 
where the dSph galaxies had apparently been destroyed.
(No significant population of gas-rich dwarf irregulars were detected.)
Both sets of authors suggested that the
faintest dwarfs in the cluster core may have been tidally destroyed
or accreted.  However, Bernstein \etal\ 1995 identified dwarf galaxies
in the Coma core down to almost $M_R{\,=\,}{-}11$, although by this point the
great majority of the objects were GCs, even $\sim\,5^\prime$ 
($\sim\,$100~\hkpc) from the central galaxy.

Numerical simulations by Bassino \etal\ (1994)
showed that the only surviving remants of dwarf satellite galaxies in the
halo of an M87-like giant galaxy were the dense centers of nucleated
dE galaxies.  (Non-nucleated dE galaxies were destroyed entirely after
a few orbits.)  For nuclei with luminosities comparable to GCs to
survive, they need to be about as dense as GCs, i.e., they would
at that point essentially {\it be} GCs.  Thus, we do not believe
very faint dwarf satellite galaxies contaminate our variance measurement;
the consistency of our results for the GCLFs presented in 
\S\ref{ssec:widths} support this conclusion.
In the following section, we tabulate our measurements of $P_0$
and $P_{GC}$ along with the measured point source number densities.

\subsection{Constraining \sn\ and $\sigma$}\label{ssec:constrain}

Table~\ref{tab:datab} summarizes the point source and fluctuation
measurements, before and after background corrections.
For each annular region of each galaxy, the table lists:
bright cutoff magnitude $m_b$ of the point source counts;
corrected number of point sources $N_{ps}$ (arcmin$^{-2}$) fainter than $m_b$ 
but brighter than $m_c$; corrected number of GCs $N_{GC}$ (arcmin$^{-2}$)
over the same magnitude range following background subtraction;
faint end cutoff magnitude $m_c$;
fitted fluctuation power $P_0$ from objects fainter than $m_c$,
in units of $10^3\,$($e^-$/pixel)$^2$;
background-subtracted power $P_{GC}$ due to GCs fainter than $m_c$,
also in $10^3\,$($e^-$/pixel)$^2$.  The annular regions are defined
as follows: c1, 32--64 pix; c2, 64--128 pix; c3, 128--256 pix; and
c4, 256--512 pix.

In order to use these measurements to derive the total number
of GCs around each program galaxy, and from that, the
galaxy's specific frequency \sn, we must first estimate the
value of the GCLF turnover $m^0$ in each galaxy.
The value of $\mo_V$ in Virgo is 
$\mo_V(\rm Virgo){\,=\,}23.75\pm0.05$
(Whitmore \etal\ 1995; Secker \& Harris 1993).  Transforming to 
the $R$~band increases the uncertainty to $\sim\,$0.07 mag.
Although Virgo is too near to have been included in the Abell
Catalog, its galaxy density is equivalent to that of a richness class~1
Abell cluster (Girardi \etal\ 1995), and it is therefore a valid
calibrating object.
The CMB velocity of Virgo is 
$v_{CMB}{=\,}1310\pm75$~\kms\ (see \S\ref{ssec:qual}),
contributing an uncertainty of 0.12~mag to the calibration.
Combining the two sources of uncertainty in quadrature yields an estimated 
error in the $m^0$ calibration of $\sim\,$0.15~mag.  This systematic
uncertainty will be left out of the error calculations for now, but we will 
discuss its effects later.

We adopt a value of
0.2~mag as a best estimate of the intrinsic dispersion in \Mo\ among giant
ellipticals in clusters (Harris 1996).
It is also necessary to consider the possibility of Abell cluster 
radial peculiar velocities 
with respect to the CMB frame; these will have an effect on both $m^0$ and
the estimated galaxy luminosity.  There is a substantial literature on
this subject (e.g. Aaronson \etal\ 1989; Huchra \etal\ 1990;
Postman \etal\ 1992; Nichol \etal\ 1992; Zucca \etal\ 1993;
Bahcall \& Oh 1996).  We  will adopt a compromise value of 400 \kms\ 
for the rms radial peculiar velocities of Abell clusters in the CMB frame.
Thus, the random uncertainty in the estimated \mo\ values for our
sample galaxies is
\begin{eqnarray}
	\label{eq:m0unc} 
	\delta \mo \, & = & \pm\, \sqrt{0.20^2 + \left({5 \over \ln 10} 
	 	\times {400 \over cz_{CMB} } \right)^2} 
		\nonumber \\ \noalign{\vskip 6pt}
		&\approx& \pm\,0.26\; \hbox{mag, at 5,000 \kms}
	     \\ &\approx& \pm\,0.22\; \hbox{mag, at 10,000 \kms}\, .
	\nonumber
\end{eqnarray}

We follow the identical $\chi^2$ minimization procedure as \cite{bt95}.
For each of the four radial regions of each galaxy, we use the corrected
variances and counts from Table~\ref{tab:datab} to calculate
the $\chi^2$ values for a grid of points in the $N_0$-$\sigma$ plane:
\begin{equation}
\chi^2 = \left( N_0 - N_0^{flu}(\sigma) 
	\over \delta N_0^{flu}\right)^2
  + \; \left( N_0 - N_0^{cnt}(\sigma)
	\over \delta N_0^{cnt}\right)^2\, ,
	\label{eq:c2chi2} 
\end{equation}
where $N_0$ is the magnitude-integrated 
surface density of GCs and $\sigma$ is the Gaussian width of the GCLF;
$N_0^{flu}(\sigma)$ and $N_0^{cnt}(\sigma)$ are the values of $N_0$
determined from the fluctuations and counts, respectively, at a specific
value of $\sigma$, and the denominators represent the uncertainties in these
quantities. Thus, the grid points represent  model GCLFs of the same \mo,
determined for each cluster according to its CMB velocity and
Eq.~(\ref{eq:m0ofz}), but differing normalizations and widths.
Blakeslee (1997b) showed the $\chi^2$ probability contours for each galaxy.
Standard $K$-corrections (Schneider \etal\
1983) were applied to the \mo\ estimates before calculating 
$N_0^{flu}$ and $N_0^{cnt}$, but these corrections
are small, amounting to just 0.03~mag in $R$ and 0.025~mag in $I$
at the limit of the survey; the same corrections were made to the galaxy
light, as it is very roughly the same color.

\section{RESULTS}

\subsection{GCLF Widths}\label{ssec:widths}

To determine final $\sigma$ values for each galaxy, we averaged the
results from the $\chi^2$ minimizations for the useful regions of each
galaxy, then varied \mo\ according to Eq.~(\ref{eq:m0unc}),
re-minimized $\chi^2$, and re-averaged.
The uncertainty in $\sigma$ due to this (random) uncertainty in \mo\ was
added in quadrature to the internal error from the $\chi^2$ minimization.
These two sources of uncertainty were usually comparable in size.

\begin{figure}\epsscale{0.7}\plotone{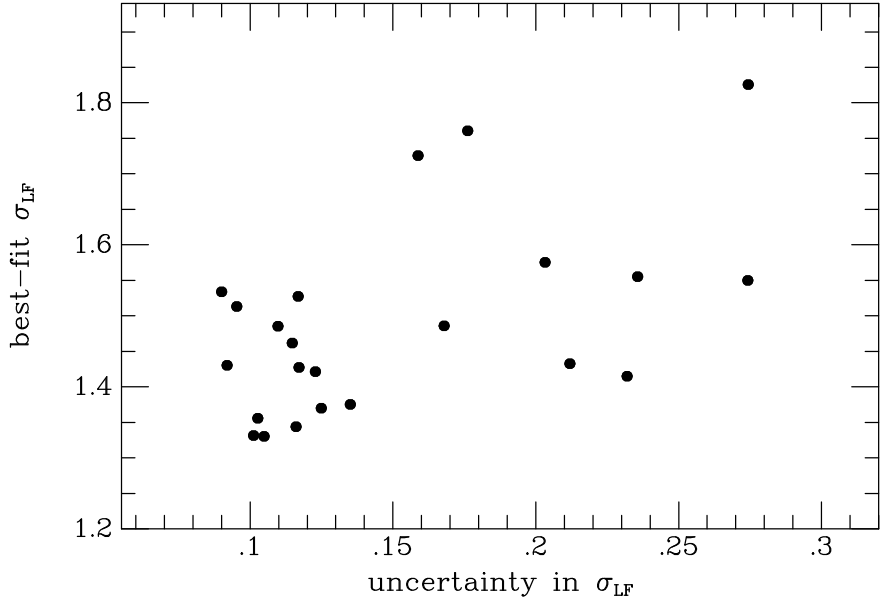}
\caption[fig03.ps]{ 
The derived GCLF width $\sigma$ in the CMB frame
is plotted against its uncertainty, showing that the more
uncertain values tend to be biased high.  See text for details.
\label{fig:sigunc_cmb}}
\end{figure}

Figure~\ref{fig:sigunc_cmb} plots the derived $\sigma$ values from this
analysis against their uncertainties.  The two Coma galaxies from \cite{bt95} 
are included in the figure.  (For values with asymmetric errorbars,
the average of the errors is used.)  The results cluster near 
$\sigma{=}1.4$~mag when the uncertainty is small, but drift higher
when the uncertainty becomes large.  Table~\ref{tab:snsig_cmb} in
the following section lists the individual values.  The overall weighted mean 
is $\langle \sigma \rangle {\,=\,} 1.45\pm0.03$~mag, the median is
1.46 mag, and the unweighted mean is 1.49~mag with a dispersion
of 0.13~mag.  (Without the \cite{bt95}  galaxies, the weighted mean,
median, and unweighted values become 1.46, 1.49, and 1.50~mag,
respectively, with the same dispersion.) 

However, if we exclude the values of $\sigma$ with uncertainties
greater than 0.15~mag, where the upward bias appears to set in, the
weighted mean is 
$\langle \sigma \rangle {\,=\,} 1.43\pm0.03$~mag, and the median and
unweighted mean are both 1.42~mag, with a 0.07~mag dispersion. 
(Excluding the \cite{bt95}  galaxies now
makes no difference in these numbers.)  It is not possible to eliminate
this bias toward larger $\sigma$ by altering the GCLF \mo\ calibration in
any reasonable way. (For instance, adopting the Baum \etal\ [1997] calibration 
makes no difference in the size of the apparent bias.)
We tentatively conclude that the GCLF width is the same for the whole sample.
However, the more poorly determined values are biased towards
larger $\sigma$, which causes one to wonder if there might also be some
bias remaining in the mean of the low-error results. 
We consult the literature for insight into this question, then offer an
explanation as to the most likely cause of the bias.

The primary GCLF calibrator for this analysis was M87, which is known to 
have $\sigma{=}1.40\pm0.06$~mag from HST measurements (Whitmore \etal\ 1995).
Other well-measured GCLFs which provide reasonable comparisons 
for our sample are those of the Virgo BCG NGC~4472,
$\sigma{=}1.47\pm0.10$~mag (Secker \& Harris 1993); the Fornax cD NGC~1399,
$\sigma{=}1.38\pm0.09$ (\cite{bt96}); the giant Virgo background
elliptical NGC~4365, $\sigma{=}1.41\pm0.15$ (Forbes 1996, HST
measurement); and NGC~5846, the central giant elliptical in a compact
group, $\sigma{=}1.34\pm0.06$ (Forbes \etal\ 1996b, also with HST).  Thus, we
expect the galaxies in this sample to have $\sigma$ near 1.40~mag, as we
found above for the well-constrained measurements.
(Note that BT95 found $\sigma{=}1.70$~mag for M87, but that was
with $m^0$ fixed to be 0.45~mag too faint, 
based on the McLaughlin \etal\ [1994] results; setting $m^0$ to the
proper value yielded a $\sigma$ of 1.45~mag). 

In recent years, other functional forms besides Gaussians have been
explored for fitting the GCLF.  The most commonly used alternative
is the ``$t_5$'' distribution proposed by Secker (1992), but intersecting
exponentials (Baum \etal\ 1995), tilted hyperbolic functions
(Baum \etal\ 1997), and Gauss-Hermite expansions 
(Abraham \& van den Bergh 1995) have also been tried.
What these alternate forms have in common is that their tails 
are broader than Gaussian, i.e., they are all ``wingy'', and they were all
motivated by the smaller GCSs (200--300 members) of the Milky Way and M31.
The GCLFs of large ellipticals with many thousand GCs, such as M87 
(Whitmore \etal\ 1995) and NGC~1399 (\cite{bt96}), are usually quite
Gaussian.  Thus, we would expect the GCLFs
of our program galaxies to be Gaussian.  As Abraham \& van den Bergh
conclude, ``for most applications, a simple Gaussian description is an 
adequate representation of the data.''

However, the galaxies in our sample with poorly constrained and
obviously biased measured values of $\sigma$ are the ones which are either
among the smallest and relatively poorest in GCs (A634-1, A1016-1) or,
in the case of A2162-1, suffered from poor seeing so that only the
very brightest members far out in the tail contributed to the direct
counts.  If the counts are enhanced relative to the Gaussian model used
for the $\chi^2$ minimization, the derived $\sigma$ will be too large.
This was pointed out by \cite{bt95} and graphically illustrated in
Figure~2-4 of Blakeslee (1997b).  We believe this to
be the most likely explanation for the bias in the under-constrained
widths.  A direct comparison in the following section of \sn\ derived
from counts and fluctuations with fixed $\sigma$ supports this view.
The galaxies for which the counts penetrated to greater depths
along the GCLF, or have no excess of bright GCs relative to a Gaussian,
will not be similarly biased.  The fact that the median $\sigma$ of
the low-error measurements is 1.42~mag, virtually identical to the M87
calibrating GCLF value of 1.40~mag, supports this view.  Moreover, it
indicates that the working hypothesis of a universal GCLF for central
galaxies in rich clusters has been vindicated.  

To summarize, we have measured the GCLF 
widths of the Abell cluster galaxies
in this sample.  Although the under-constrained values are biased
high, the remainder appear to be accurate.  We find no evidence for any
intrinsic dispersion among these better measured widths, and their
average is very close to the value for M87, whose turnover magnitude 
\mo\ and CMB velocity were used to estimate \mo\ for the program galaxies,
thereby allowing us to derive the GCLF widths.
This result provides a valuable consistency check in support of the assumption
of a universal GCLF for bright ellipticals in Abell clusters.
In the previous section, we estimated the systematic uncertainty in
the \mo\ calibration to be $\pm0.15$ mag, mainly due to the uncertainty in
the CMB velocity of Virgo.  Varying the estimates of \mo\ by this amount
has the effect of changing the individual GCLF widths by 0.04--0.06 mag.
Thus, we conclude that among the GCLF width measurements
with internal uncertainties smaller than 0.15~mag, the mean is
$\langle \sigma \rangle {\,=\,} 1.43\pm0.03\pm0.05$ mag, where the
first errorbar reflects the internal error and the second indicates
the systematic uncertainty.  We note that if we had set the
Virgo CMB velocity using the Baum \etal\ (1997) value of the relative 
Coma-Virgo distance, the derived values of $\sigma$ would be 
larger by 0.10-0.15 mag, and $\langle \sigma \rangle$
would increase by 0.12~mag.  A discrepancy of this size
would be difficult to resolve.

Because we find no evidence of significant differences in GCLF $\sigma$ between
M87 and the sample galaxies or among the sample galaxies themselves, 
we report specific frequencies in the following section assuming the HST
M87 value of $\sigma{\,=\,}1.40$~mag.
For an intrinsic dispersion in this value, we assume 
$\pm0.05$~mag, consistent with our low-error results 
(which are in fact consistent with no intrinsic dispersion)
and with the scatter in the measured widths of nearby dominant ellipticals.
This approach was chosen because the derived
\sn\ is strongly anticorrelated with the value of~$\sigma$.
In this way, we hope to avoid biased intercomparisons among the \sn\ results
for these galaxies.

\subsection{Specific Frequencies}\label{ssec:sn}

If $N_{GC}$ is the total number of GCs in a galaxy and $M_V$ is its
absolute $V$ magnitude, the specific frequency is 
$\,\sn = N_{GC}{\,\times\,}10^{\,0.4(M_V + 15)}$.
In order to calculate \sn, one must adopt a zero point for the distance
scale. Up to this point, all magnitudes have been set relative to
their observed values in Virgo, using the Coma-Virgo relative distance to
place Virgo in the CMB frame.  
We now adopt a Virgo distance modulus of 31.02 based on four Virgo
spirals with Cepheid distance moduli 
(Freedman \etal\ 1996; there is also one about 1~mag more distant). 
Our assumptions imply a Hubble constant $H_0\,{\sim\,}80$~\kms~ Mpc$^{-1}$,
but this is irrelevant for our primary goal of
measuring trends in \sn.
Transforming the \sn\ values presented below
to a different distance scale requires multiplying them all by
$10^{\,0.4\,[\,31.02 - \mM_{0}]}$, where $\mM_{0}$ is the alternative
Virgo distance modulus.  

For each radial region of each galaxy, we assumed
$\sigma{\,=\,}1.4$~mag and calculated the total GC surface
density $N_0$ from the counts of GCs brighter than
the cutoff magnitude $m_c$ and from the fluctuations resulting from
those fainter than this cutoff.  The two separate measurements were
then weighted averaged,
and all the GCs between some inner radial limit, discussed
below, and an outer radius of 40~kpc (or 32~\hkpc\ and $h{\,=\,}0.8$)
were summed; the result was then
divided by the normalized luminosity of the galaxy within the same radial
range to derive the ``metric $S_N$'' within 40~kpc.  This was repeated
for $\sigma{\,=\,}1.35$ and 1.45~mag, and then for $\sigma{\,=\,}1.4$ but
with $m^0$ varied according to Eq.~(\ref{eq:m0unc}).  (For \sn, unlike
$\sigma$, the distance uncertainty due to random cluster velocities largely
cancels, since it similarly effects both $N_0$ and the galaxy luminosity.)
The variation
in \sn\ due to the uncertainties in these GCLF parameters was then added
in quadrature to the internal errors which had been propagated from
the counts and fluctuations.  Table~\ref{tab:snsig_cmb} lists the
metric \sn\ results, with their final errors, for each galaxy.
(These were previously presented by Blakeslee [1997a,b].)
It also lists the derived values of $\sigma$ discussed in
the previous section, but these are not the ones used in calculating
the tabulated values of \sn.  

For reference, we list in Table~\ref{tab:snflucount} the specific frequency
values $S_N^{flu}$ and $S_N^{cnt}$ derived separately from the variance 
measurements and the counts, respectively, with GCLF $\sigma$ fixed at
1.40~mag.  The internal measurement errors shown in the table
include no allowance for uncertainty in the GCLF \mo\ or $\sigma$, and
are therefore uncorrelated. The value of the width $\sigma$ which would
bring $S_N^{flu}$ and $S_N^{cnt}$ into precise agreement for a given galaxy
is the one listed in Table~\ref{tab:snsig_cmb} for that galaxy.
Table~\ref{tab:snflucount} also lists the number of standard
deviations $\sigma_{dev}$ separating the two \sn\ values.  Note that the
four galaxies with $\sigma_{dev}{>\,}$2 all have $S_N^{cnt}{>}S_N^{flu}$ 
and best-fit $\sigma\ge1.53$ mag (Table~\ref{tab:snsig_cmb}), consistent
with the bias in the GCLF width measurements being
due to counts which are enhanced relative to
a Gaussian model.  However, we did not feel that this would justify throwing 
away the information provided by the counts.

Estimates of the global \sn\ involve two very uncertain extrapolations
of the GCS and galaxy profiles out to large radii.  Not only are small
uncertainties magnified by the extrapolations, but there is often
no guarantee that the profiles do not change outside the imaged
field.  Thus, we chose not to make such extrapolations.
The decision to report metric values of \sn\ (i.e., values derived within
the same physical radius around each galaxy) was made in order to avoid
a bias in the reported \sn\ with redshift.  The metric radius of 40~kpc
was chosen because it corresponds to the limit of the image for the
nearest of the sample galaxies (roughly 500~pix, or 2\farcm3, in the
1024$^2$ $R$-band images).  Finally, the inner radius varied a bit
and was set by practical constraints from the 
variable quality of the model subtraction at small radii.
In effect, the inner limit was the smallest usable radius, typically
smaller (in angle on the sky) for the more distant galaxies, but never
less than 10\arcsec, corresponding roughly to the expected GCS core 
radius in the nearest galaxies (Forbes \etal\ 1996a). 
It makes little difference, however, as such a small portion of
the GCS is involved. 
Note that since \sn\ is number per unit galaxy luminosity, if the
GCs follow the same radial distribution as the halo light (true to at least
first order), the ``metric \sn'' will actually be independent of the
chosen metric radius and will equal the global \sn. 
Henceforward, by \sn\ we will mean these metric specific frequencies, 
unless otherwise stated.

The \sn\ values listed in the table for A1656-1 (NGC~4889) and A1656-2
(NGC~4874) differ from the global values quoted in \cite{bt95} for two reasons.
First, the RC2 photometry (de Vaucouleurs \etal\ 1976) used there was in
error.  It gave a total magnitude for NGC~4874 that was substantially 
too faint, as well as a slightly too faint total magnitude for NGC~4889.
Using the photometry from the RC3
(de Vaucouleurs \etal\ 1991) yields global values of
$S_N{\,=\,}10.2\pm2.4$ for NGC 4874 and $S_N{\,=\,}6.8\pm1.8$ for NGC 4889,
closer to those shown in Table~\ref{tab:snsig_cmb}.  Second, as stated above,
the table gives the metric \sn\ values within a limited radial range,
using our own photometry of the galaxy light within that range.  As the
GC systems of those two galaxies were found to be more extended than the
halo light, it is expected that their metric \sn\ would be smaller than
their global \sn, and we see from the table that this is the case.

We conclude with some comments on the measured values of \sn\ for these
galaxies and their uncertainties.   Our results indicate 
the existence of a continuum of \sn\ values, rather than a separation
of galaxies into ``normal'' and ``high'' $S_N$ classes.
For each of the three clusters with more
than one galaxy in our sample, it is the second brightest galaxy
that has the higher \sn\ (although in the case of A539 the difference
is not significant).  These galaxies were included in the sample
because they are all giant ellipticals with luminosities similar to
those of the corresponding BCGs selected by LP.
In fact, Table~\ref{tab:bcg} shows that they
all have larger $\alpha$ parameters than the BCGs, and
comparison of their coordinates with those in Table~\ref{tab:cls} 
indicates that they are all closer to their cluster X-ray centers.
Thus, for these three clusters, the most centrally dominant galaxy
happens to be the second brightest one.
We discuss below how central location in a cluster might affect \sn.

We have attempted to keep the \sn\ errors listed in
Table~\ref{tab:snsig_cmb} both realistic and independent of each other.
However, it is possible that they may include some systematic components,
and therefore not be completely independent of each other.
If, for example,
the intrinsic (random) dispersion in $M^0$ is only $\sim\,$0.1 mag instead 
of $\sim\,$0.2 mag, then the final errors will all be similarly overestimated,
which is to say that they would contain a systematic component,
and not be what we loosely term ``independent''. 
It seems unlikely, though, that the intrinsic
dispersion in the GCLF width is less than the 0.05~mag we have adopted.
The uncertainties from the GCLF parameters dominate, being 1--3~times
as large as the uncertainties due to measurement error,
depending on the size of the GC population and the depth of the data.
The tabulated errors are offered as best estimates
of the random (i.e., non-systematic) uncertainties in the individual
\sn\ values, but due to a preference to err on the side of caution,
it is possible that these quoted random errors may be systematically
overestimated.

Finally, we discuss the effects on the \sn\ results of the
estimated $\pm0.15$ mag systematic uncertainty in the $m^0$ calibration.
Increasing the individual \mo\ estimates by 0.15~mag has the effect
of increasing the derived \sn\ values by $19\pm3\,$\%;
decreasing \mo\ systematically by the same amount decreases the
derived \sn\ values by $15\pm2\,$\%.
Thus, the ``top'' of the \sn\ continuum for this sample
may move upward from 9.3 to 11.1,
as the ``bottom'' moves from 3 to 3.6, or the top and bottom
may decrease together to 7.9 and 2.5, respectively.
These estimates assume that the galaxy luminosities are held fixed;
varying them along with \mo\ decreases the leeway to just a few percent,
but the relative uncertainties in the Virgo distance modulus (which sets
the zero point for the luminosities) and CMB velocity (which sets \mo\
for each cluster with respect to Virgo) are similar in size.~~

\subsection{Abell Cluster Inertial Frame Results}

The values of \sn\ and GCLF width $\sigma$ reported and discussed
in previous sections were all calculated in the CMB frame.
However, there is another relevant
frame which might also have been adopted, the Abell Cluster Inertial
(ACI) frame.  The ACI frame is defined to be at rest with respect to the
large-scale Abell cluster bulk flow discovered by LP using the 
$L_m$-$\alpha$ distance indicator.  

In fact, we did the analysis twice simultaneously,
using both CMB and ACI frame velocities to estimate \mo\ and galaxy
luminosities.  The same distance zero point was used in 
both analyses; however, from one reference frame to the other,
the relative distances between the
galaxies change as a function of galactic coordinates,
according to the transformation given by LP.
Table~\ref{tab:snsig_aci} lists the \sn\ and $\sigma$ results of our
analysis in the ACI frame.  The quantities were derived in an identical
manner to those shown in Table~\ref{tab:snsig_cmb}, but the ACI frame
velocities were used instead of CMB velocities.  The \sn\ values are
generally higher.  This is a consequence of the fact that the
Coma-Virgo distance is fixed, but the velocity of Coma is nearly 10\%
smaller in the ACI frame (see Table~\ref{tab:cls}).  

\begin{figure}\epsscale{0.7}\plotone{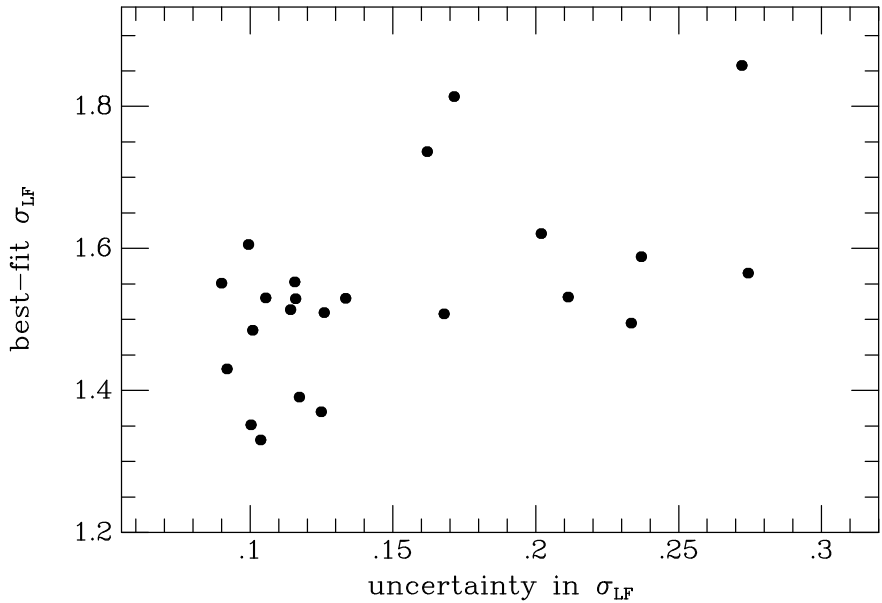}
\caption[fig04.ps]{ 
The derived GCLF width $\sigma$ in the ACI frame is plotted
against its uncertainty.  Again, the more uncertain values
tend to be high, but here the median of the well-determined 
values is also high compared to the CMB frame results.
\label{fig:sigunc_aci}}
\end{figure}

Among the galaxies in clusters without fixed distances
(i.e., everything but Coma), the fractional scatter in $S_N$ increases by 5\% 
(and the absolute scatter increases by 11\%)
when the ACI frame is used instead of the CMB frame.
In addition, the mean and median GCLF $\sigma$ values among those with
errors smaller than 0.15~mag are now $\langle \sigma \rangle{\,=\,}1.48$~mag
and $\sigma_{med}{\,=\,}1.51$~mag, respectively; dropping the
``fixed point'' Coma galaxies, these values increase to 1.51 and 1.53~mag.
(See Figure~\ref{fig:sigunc_aci}.)
Thus, unlike the results for $\sigma$ in the CMB frame, 
which were discussed in \S\ref{ssec:widths}, we do not find good agreement
between the values of $\sigma$ in the ACI frame and the 
$\sigma{\,=\,}1.40$~mag width of the GCLF \mo\ calibrator M87.
For these reasons, we have decided to concentrate the rest of the analysis
on the results found using CMB frame velocities.

\section{$S_N$ CORRELATIONS}

The \sn\ values derived in the CMB frame are plotted below
against various properties of the host galaxy and surrounding cluster.
To prevent confusion, we remind the reader that 
the word ``cluster'', when it appears alone, means ``galaxy cluster'';
the abbreviation ``GC'' refers to ``globular cluster.''

\subsection{Galaxy Properties}

\begin{figure}\epsscale{0.85}\plotone{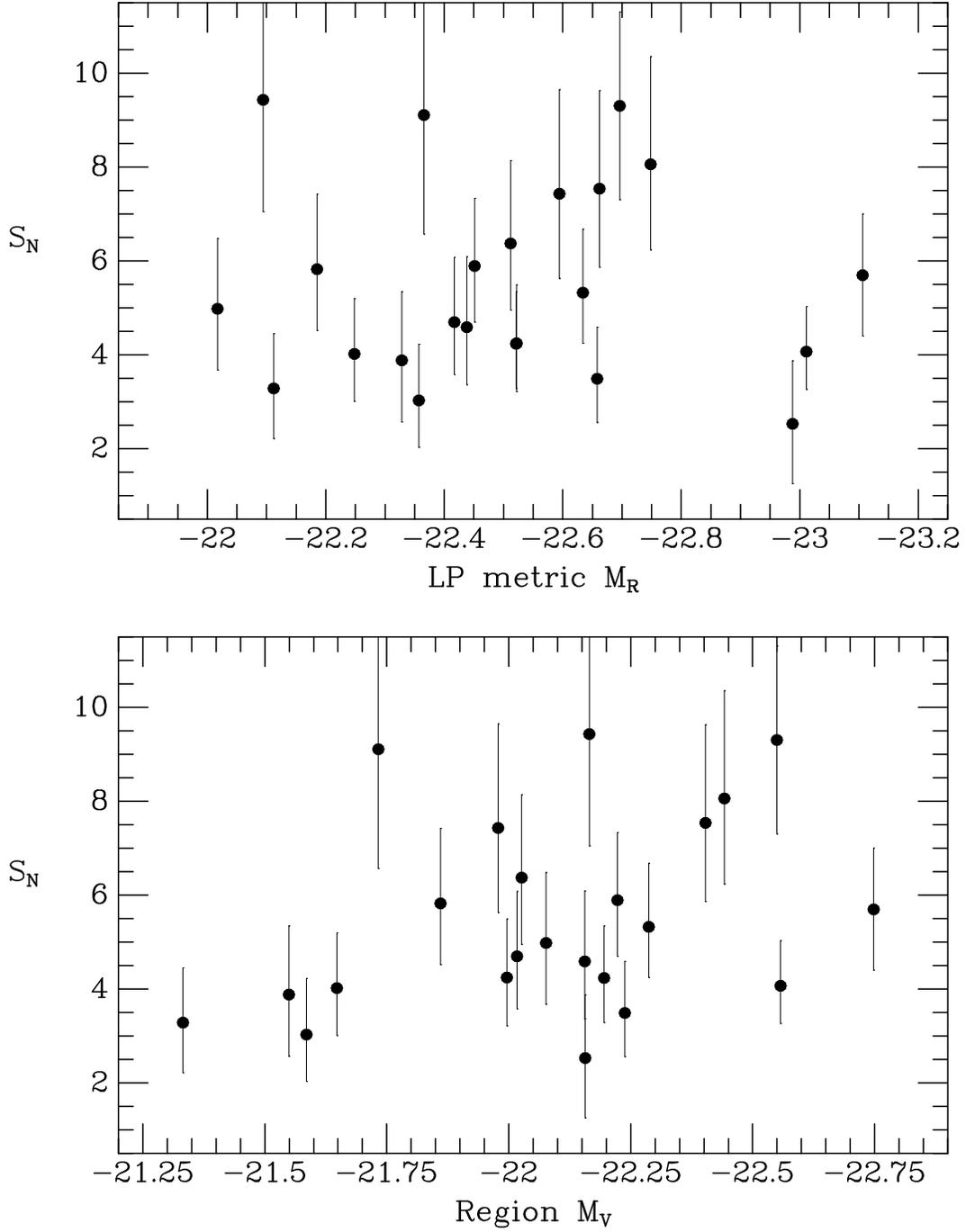}
\caption[fig05.ps]{ 
$S_N$ is plotted against $(a)$ metric absolute magnitude from Lauer \&
Postman (1994), and $(b)$ total absolute magnitude of the specific region 
of each galaxy in which $S_N$ was measured (i.e., from $\sim\,$3 to 32 \hkpc).
All quantities are calculated in the CMB frame. \label{fig:sn_mag}}
\end{figure}

The top panel of Figure~\ref{fig:sn_mag} plots the derived $S_N$ 
values against absolute $R$-band CMB metric magnitude of the galaxy from LP.
The lack of any significant correlation is understandable,
as the LP metric magnitude is calculated within 12.5~kpc, while these
$S_N$ values were derived for the total usable area of the galaxy within
40~kpc.  Therefore, in the lower panel of the figure we plot $S_N$ against
the absolute $V$~magnitude (since $S_N$ is defined relative to V magnitude)
of the region of each galaxy in which the plotted $S_N$ was derived.  
Here, there does appear to be a weak correlation, but this is based almost
entirely on the four faintest galaxies, which all have $S_N{\,\lta\,}4$.
Overall, the significance of the correlation is $\sim\,$0.95,
based on the Spearman rank-order correlation coefficient,
but excluding the four faintest galaxies causes the significance
level to drop to $\sim\,$0.40 (i.e., no correlation).
Galaxy luminosity is apparently not the driving force
behind variations in \sn\ among these galaxies. 
The panels of Figure~\ref{fig:sn_mag} include the errorbars from
Table~\ref{tab:snsig_cmb}, primarily to demonstrate their size, but
the rest of the figures will omit them for the sake of clarity (and
because of the possibility of systematically overestimated random errors,
discussed above).

\begin{figure}\epsscale{0.85}\plotone{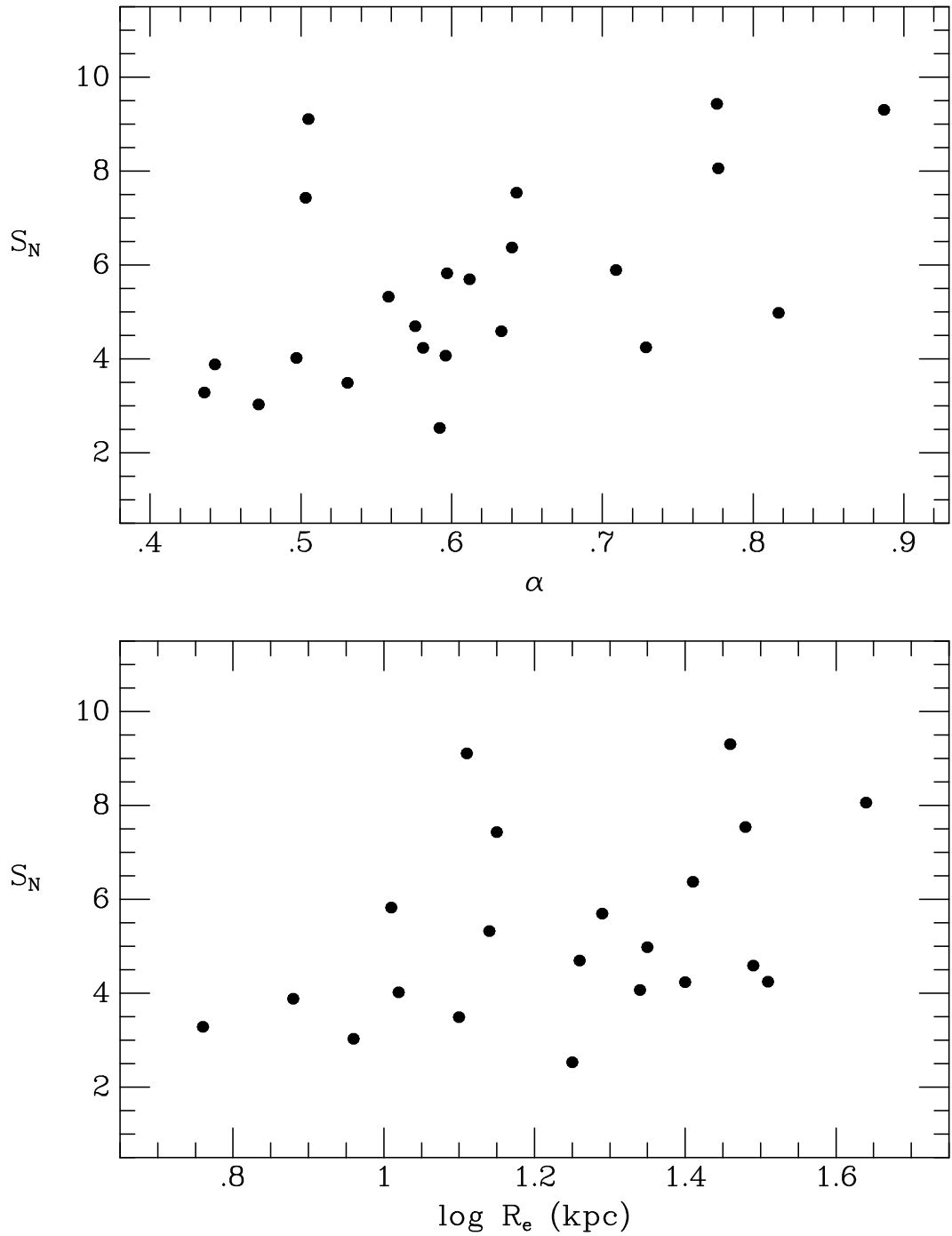}
\caption[fig06.ps]{ 
$S_N$ is plotted against two measures of the galaxy profile,
$(a)$ the structure parameter $\alpha$ and $(b)$ the logarithm of
the effective radius $R_e$.  Errorbars have been omitted for the sake of
clarity.
\label{fig:sn_ext}}
\end{figure}

High values of \sn\ have been associated with the phenomenon of extended
cD halos (e.g., Harris 1991, McLaughlin \etal\ 1994).
Figure~\ref{fig:sn_ext} shows $S_N$ plotted against two measures of galaxy
extent, the profile structure parameter~$\alpha$ (defined in \S\ref{sec:sample}),
as listed by LP for the CMB frame, and the effective radius $R_e$ of the best
fitting \roneq-law, taken from Graham \etal\ (1996), who tabulated this
quantity for the BCGs in the CMB frame with $H_0{\,=\,}80$ \kms~Mpc$^{-1}$.
Where available, the effective radii for the non-BCGs were taken from 
the RC3 and brought into consistency with 
Graham \etal\ via the effective radius of the BCG, tabulated by both sources.
Both panels exhibit a general upward trend. 
Since $\alpha$ measures the slope of
the galaxy profile at fairly small radius, $R_e$ is probably a better measure 
of overall halo extent, but even with two apparent outliers,
the top panel has less scatter. 
The biggest outlier in the $\alpha$ plot is A539-1 (with 
$\alpha{\,\sim\,}0.5$, $\sn{\,=\,}9.1$); this
is not the dominant galaxy in A539,
judged by position relative to the cluster X-ray/dynamical center,
although it does have many GCs.
The other ``outlier'' is A2162-1 ($\alpha{\,\sim\,}0.5$,
$\sn{\,=\,}7.4$), a cD whose extended halo only
sets in at larger radius; it begins to move into line with
the more extended galaxies in the lower panel.
Although both plots show considerable scatter, they do support
the view that \sn\ and galaxy extent are somehow associated.

\begin{figure}\epsscale{0.9} \plotone{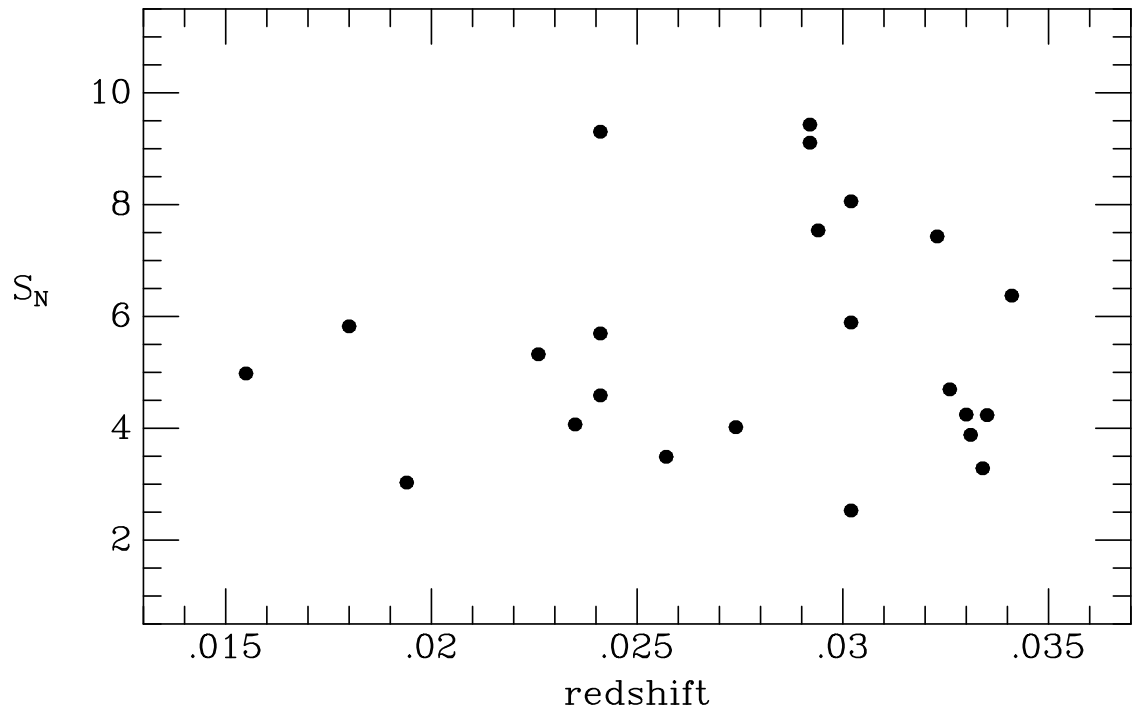}
\caption[fig07.ps]{ 
$S_N$ is plotted against the cluster redshift in the CMB frame.
Though this is the redshift used in deriving the $S_N$ values,
there is a gratifying lack of correlation.
\label{fig:sn_red}}
\end{figure}

It is always advisable to look for systematic effects in final results;
therefore in Figure~\ref{fig:sn_red} we have plotted our derived \sn\ values
against cluster redshift.   These redshifts were the basis for our \mo\
estimates and the galaxy luminosity calculations, but as the figure shows,
they do not correlate with \sn.

\subsection{Cluster Properties}\label{ssec:corr}

We now consider possible correlations of \sn\ with properties of the
clusters in which the galaxies reside.  There are many different
parameters to explore, including density, dynamics, morphology,
and X-ray properties,  any of which could have an effect on the
GCS of the central galaxy in the cluster.
Correlations of $S_N$ with some of the properties reflecting
cluster mass or density have been discussed by Blakeslee (1997a), but
here we provide a more complete discussion.  We begin with the dynamics.

In collecting cluster velocity dispersions from the literature, one
must take care to ensure that the final set is a fairly homogenous one.
Girardi et al. (1993) found that the various methods of estimating 
dispersions give consistent results as long as the number of cluster 
members with measured redshifts exceeds about 20; less sophisticated, 
or ``robust,'' analyses often yield erroneous results with fewer than 15
measured redshifts.   This collection of dispersions, listed in 
Table~\ref{tab:cls}, started with the
``robust dispersions'' determined by Girardi et al. (1993) from the 
data sets of Zabludoff et al.\ (1990) and Yahil \& Vidal (1977)
for cluster galaxies within $1.5\,h^{-1}$~Mpc (the Abell radius, $r_A$)
of the cluster centers.
These dispersions have been superseded by the more recent measurements
of Beers \etal\ (1991) for A569 and Zabludoff \etal\ (1993) for A1185,
A1367, A1656, and A2199, both of whom likewise reported dispersions
within $1\,r_A$.  To these have been added the dispersions
for A2634 and A2666 from Scodeggio \etal\ (1995), who conducted a
detailed study of these possibly interacting clusters.  Because A2634 appears
to be merging with a group of spirals at larger radius, and the smaller 
A2666 would otherwise be heavily contaminated by nearby A2634 galaxies, 
the dispersions within half an Abell radius have been used; these
should more accurately reflect the central potentials of these clusters.
All of these dispersions are based on at least 22 cluster members with
measured velocities (after outliers have been rejected). Finally,
Struble \& Rood (1991) list dispersions based on $>$20~member redshifts
for three other clusters in the BCG sample, so these have been used 
as well.

\begin{figure} \epsscale{0.85} \plotone{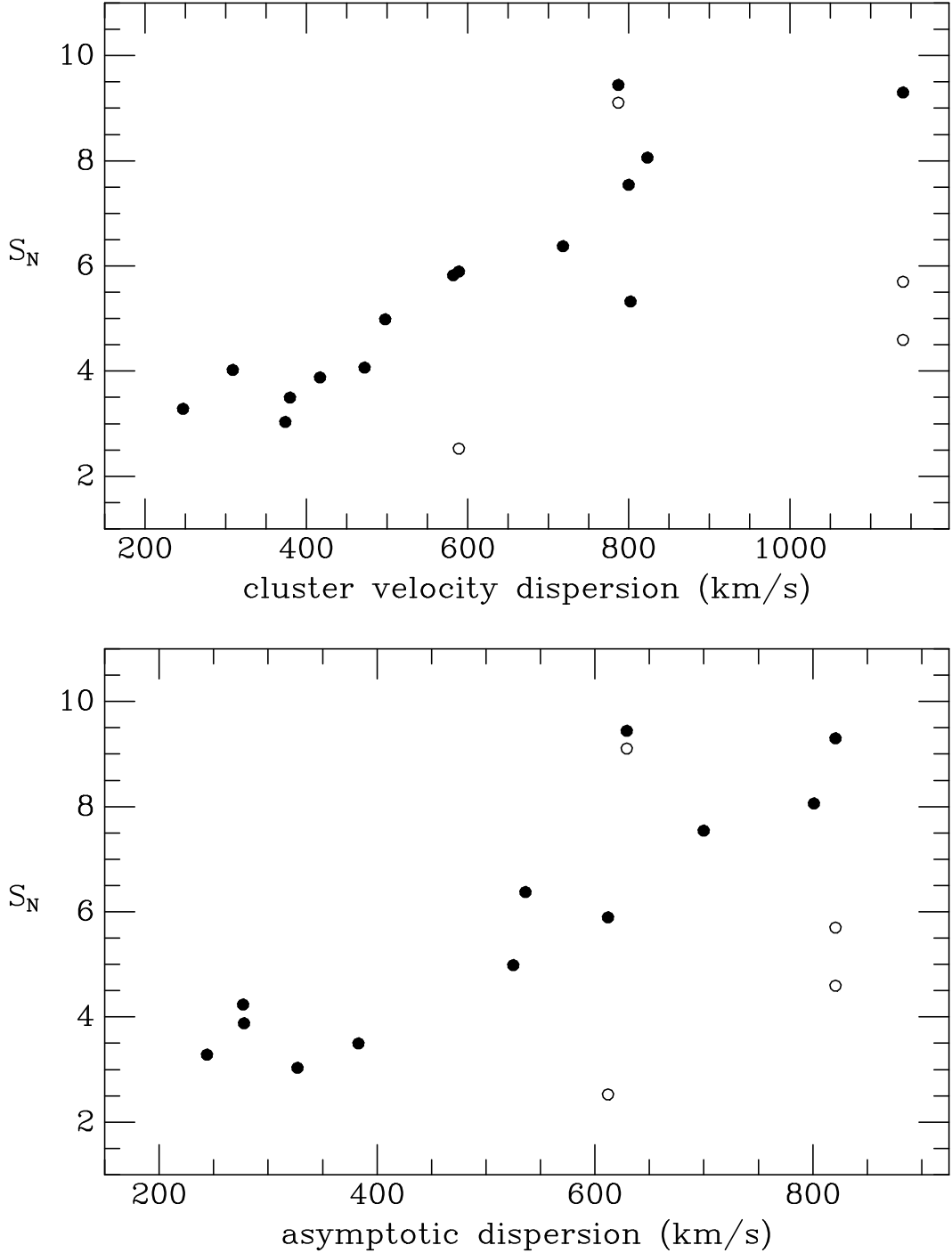}
\caption[fig08.ps]{ 
The correlation between $S_N$ and cluster central velocity dispersion.
In $(a)$, central dispersions have been collected from various sources
in the literature. 
In $(b)$  the asymptotic velocity dispersions reported by Fadda \etal\
(1996) are used.
Open symbols represent the ``secondary'' galaxies in clusters with more
than one member in the present sample
(see text for details); filled circles are rest of the sample.
\label{fig:sn_disp}}
\end{figure}

Figure~\ref{fig:sn_disp}$a$ shows the resulting plot of $S_N$ against cluster
velocity dispersion.  The tight correlation evident in the figure indicates
that bright galaxies in regions of higher dispersion, and thus deeper potential
wells, have more GCs per unit luminosity. In this and most of the 
following figures, the more central galaxies (judged by the X-ray center)
in the clusters with multiple members in this sample are the ones shown
as filled symbols, while the less central ones are shown as open symbols,
since they often deviate from the correlations.
The non-central galaxies will be referred to as ``secondary'',
although three of the four are among the BCGs selected by LP,
with the other one being
NGC~4839, the third ranked galaxy in Coma.
Correlations among cluster richness, dispersion, and
X-ray temperature are well established (e.g., Bahcall 1981;
Mushotzsky 1984; Edge \& Stewart 1991;
Lubin \& Bahcall 1993; Girardi et al. 1993, 1996) and were to be
expected as different measures of the cluster potential
(assuming galaxies trace mass).  Now we see a clear relationship between
one of these quantities and something apparently
unrelated, the GC specific frequency of the central bright
galaxy in the cluster.

Recently, Girardi and collaborators have been studying cluster
velocity dispersion profiles (Fadda \etal\ 1996; Girardi \etal\ 1996),
reporting the asymptotic values at large radii.
They argue that ``asymptotic dispersions'' are
less affected by small-scale velocity anisotropies, and thus more
indicative of the depth of the cluster potential as a whole.
Because the profiles are usually peaked in the center
before flattening further out, the asymptotic dispersions are 
generally lower than the central dispersions employed above.
(These authors also cut the profiles of A2634 and A2666 off at
smaller radii, however.)
For this reason, we separately plot the asymptotic cluster
velocity dispersions, as listed by Fadda \etal, against $S_N$ in
Figure~\ref{fig:sn_disp}$b$.  The correlation is again very strong.

\begin{figure} \epsscale{0.85}\plotone{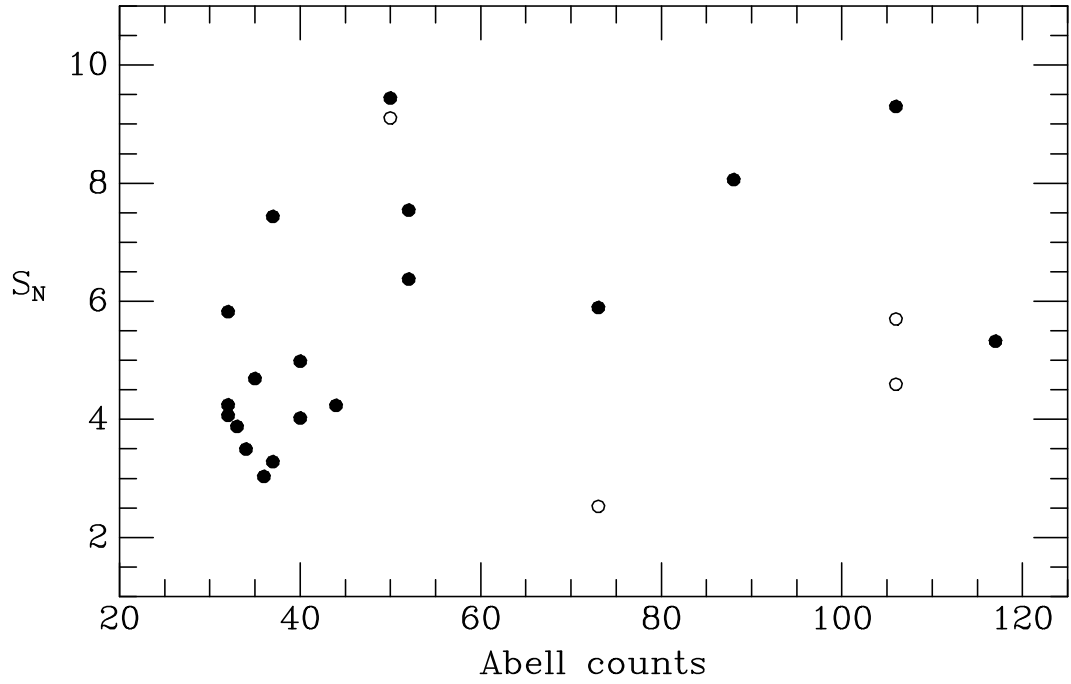}
\caption[fig09.ps]{ 
$S_N$ is plotted against Abell counts, a measure of overall cluster richness.
Symbols are as in Figure~\ref{fig:sn_disp}. 
\label{fig:sn_abell}}
\end{figure}

Figure~\ref{fig:sn_abell} shows that the correlation between $S_N$ and
galaxy density as measured by Abell galaxy counts
(Struble \& Rood 1987) is much weaker.
The high formal significance of the correlation
is due to the fact that the lowest \sn\ central galaxies are all 
in relatively poor clusters; excluding these makes the correlation 
marginal.  However, the uncertainties in the Abell counts are large 
and not well-known.  The number density of galaxies drops steeply
out to the $1\,r_A$ limit, resulting in a large amount of background
contaminaton, and making the counts a questionable measure of central
density (Beers \& Tonry~1986).~~
Bahcall (1981) showed that a better measure of the cluster density was
provided by background-corrected counts of bright galaxies within $r_A/2$.
Those counts correlated better with cluster dispersion and
X-ray luminosity than did the Abell counts.  However, the present sample
has very little overlap with the Bahcall sample; thus, we performed our
own counts of bright galaxies within an even smaller radius, set by the
size of the CCD field, making background correction completely unnecessary.

\begin{figure}\epsscale{0.85} \plotone{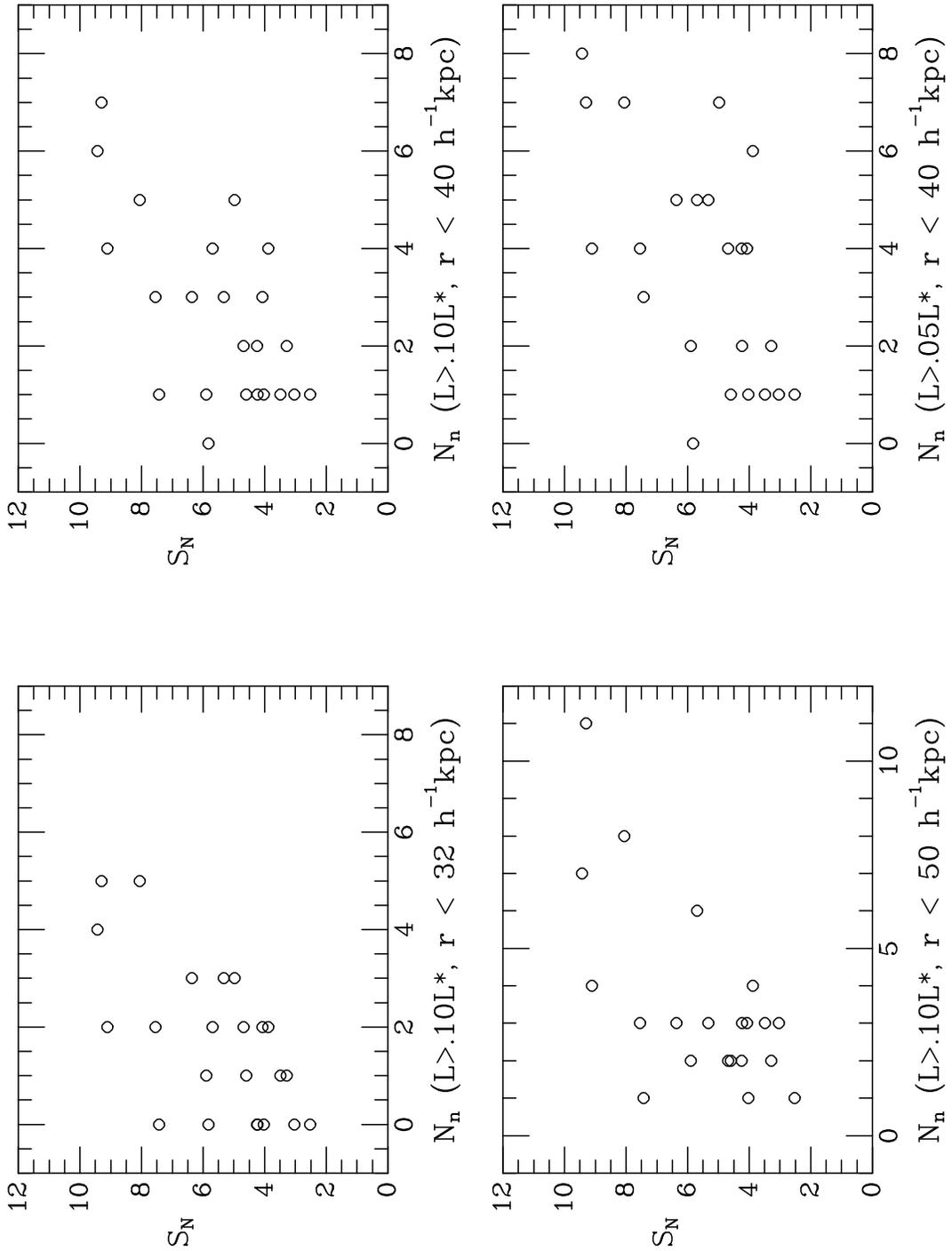}
\caption[fig10.ps]{
$S_N$ is plotted against the total number of neighboring galaxies
brighter than $0.1 L^*$ within 32, 40, and 50~\hkpc\ of the galaxy
center, and against the number of neighbors brighter than
$0.05 L^*$ within 40~\hkpc.\label{fig:jbcounts}}
\end{figure}

We counted all galaxies 
brighter than 0.05 and 0.1~$L^*$ within several
different radial distances from the program galaxy centers,
where $L^*$ is the characteristic luminosity of the Schechter (1976) function.
The value of $L^*$ was taken from Lin \etal\ (1996) and transformed from
the $r$-band to our photometric bands following Schneider \etal\ (1983).
The results of our counts are reported in
Table~\ref{tab:jbcounts} which lists the number of neighbors $N_n$ brighter
than 0.1~$L^*$ within 32, 40, and 50~\hkpc\ and the number brighter than
0.05~$L^*$ within 32 and 40~\hkpc. Figure~\hbox{\ref{fig:jbcounts}}
plots the counts against \sn.  We are hampered by
small number statistics, but it is clear that the galaxies surrounded by
more neighbors, or in regions of greater galaxy density, tend to have 
higher \sn.  Again a relationship is suggested
between central location within the cluster and \sn,
assuming that the increase $N_n$ is the result of the galaxy's 
being closer to the cluster dynamical center.  

\begin{figure}\epsscale{0.85}\plotone{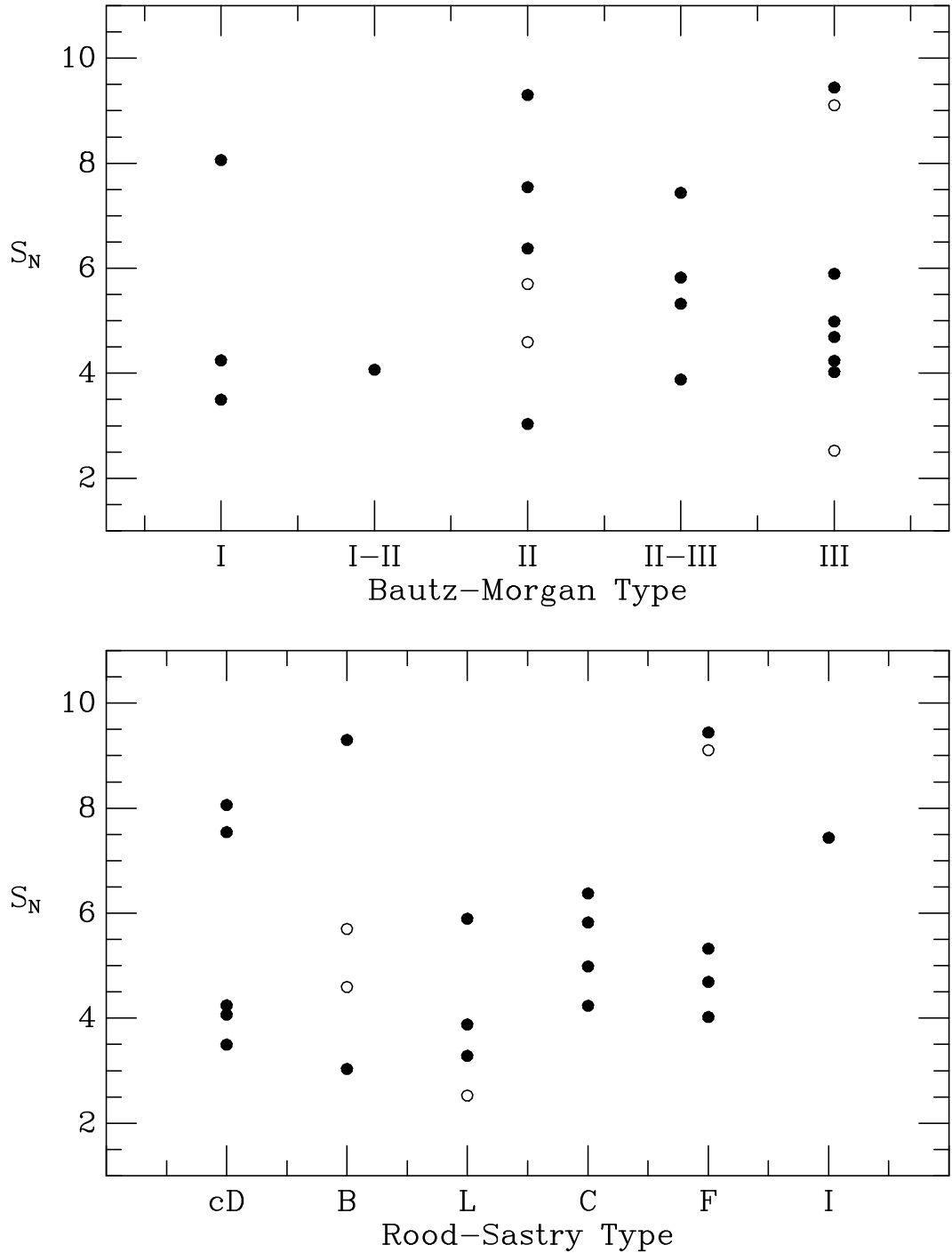}
\caption[fig11.ps]{ 
$S_N$ is plotted against morphological type in the Bautz-Morgan
and Rood-Sastry classification systems.
Unlike previous investigations based on many fewer clusters, we find no
evidence of any correlation in these data. 
Here and in the following figures, symbols are as in Figure~\ref{fig:sn_disp}.
\label{fig:sn_morph}}
\end{figure}

Moving on to morphology, we plot \sn\ against Bautz-Morgan type
(from Abell \etal\ 1989) and Rood-Sastry type (Struble \& Rood 1987) in
Figure~\ref{fig:sn_morph}.   There have been reports based on smaller
data sets that central galaxy \sn\ correlates with BM type (McLaughlin,
Harris, \& Hanes 1993, 1994; Harris, Pritchet, \& McClure 1995),
in the sense of the
``later'' types (II-III, III) having higher \sn\ central galaxies.  The
classification in this system is based on the relative dominance of the
central galaxy in the cluster, with BM~I clusters being dominated by single
giant galaxies, and BM~III cluster having no clearly dominant galaxy.
McLaughlin \etal\ (1993) suggested that the apparent
anti-correlation of \sn\ with central galaxy dominance was due to dynamical
evolution, with the central giants in the BM~I and I-II clusters having
diluted their originally high-\sn\ down to lower levels through repeated
mergers with other galaxies in the cluster.  It was an
interesting suggestion, but we do see not see a correlation of \sn\ with
BM type in this larger, more homogenous data set; thus, we find no
evidence that such \sn\ dilution is taking place.

Rood-Sastry type is also usually thought of in evolutionary terms, with the
evolutionary state becoming more advanced along the following morphological
sequence: Irregular, Flattened, Core-dominated, Linear, Binary, and cD.
Again, our immediate reaction is that no correlation exists
between central galaxy \sn\ and RS type.
However, Schombert \& West (1990) suggested that this
classifications system reflected dynamical evolution only up to the L~class,
with the B and cD classes (clusters dominated by two giant galaxies
and one giant galaxy, respectively) representing further evolution 
of the galaxy luminosity
function following cluster virialization.  This suggestion was based on
an apparent correlation of the I-F-C-L sequence with supercluster
environment.  It is conceivable that there is a systematic change of \sn\ 
along this subsequence of the RS system, but if so, it is most likely
an ancillary consequence of the dependence central galaxy \sn\ on cluster
density, seen most clearly in the correlations with velocity dispersion
(above) and X-ray properties (below).

\begin{figure}\epsscale{0.85} \plotone{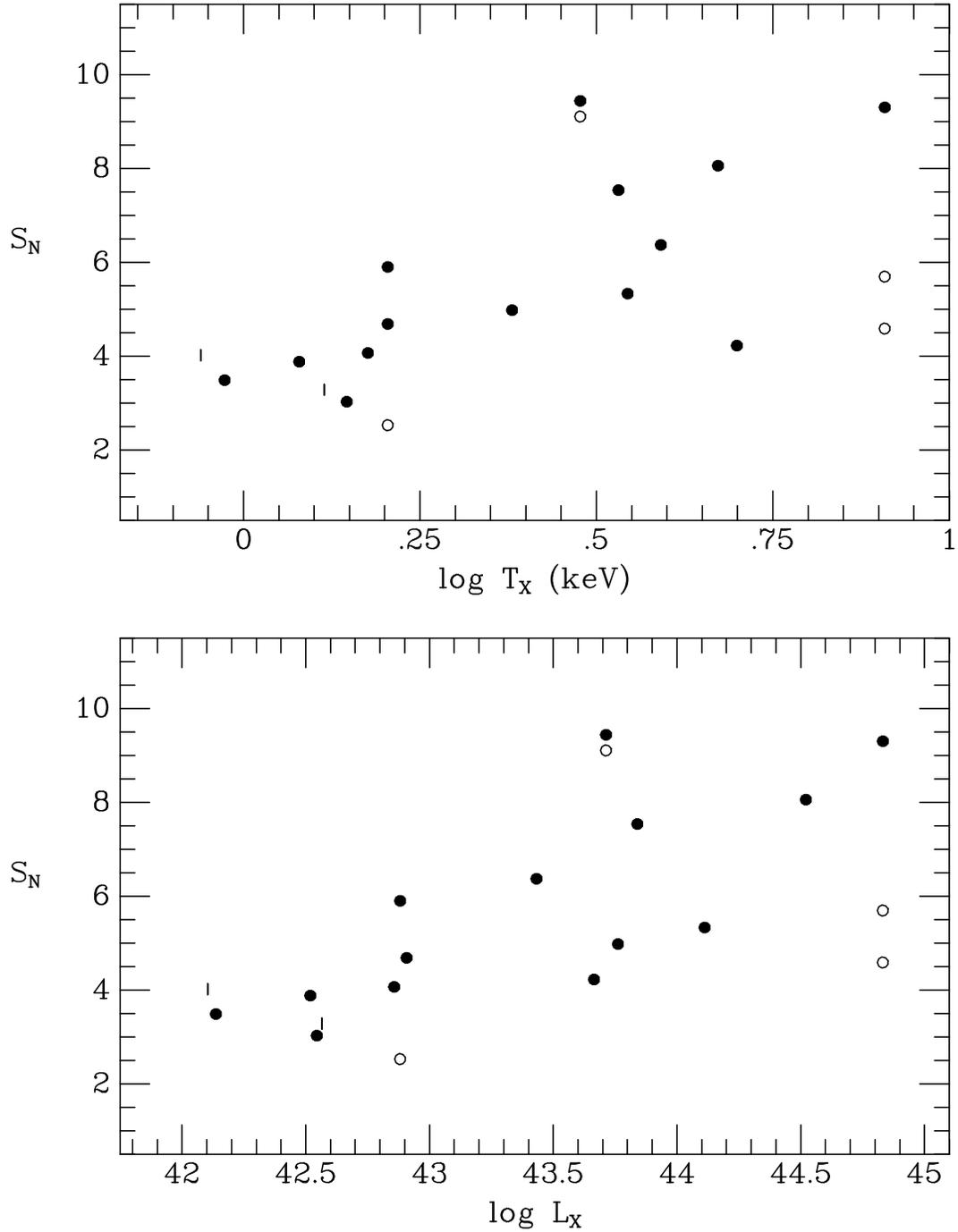}
\caption[fig12.ps]{ 
$S_N$ is plotted against $(a)$ temperature of the X-ray emitting
intra\-cluster gas, and $(b)$ total X-ray luminosity within 
1~Mpc of the cluster X-ray center
(with $H_0{\,=\,}50$, as given by Jones \& Forman 1997).
Short vertical lines represent galaxies in clusters 
with only upper limits on their X-ray emission.
\label{fig:snxprops}}
\end{figure}

From the correlation of \sn\ with cluster velocity dispersion
found in Figure~\ref{fig:sn_disp}, and the known relation between
cluster velocity dispersion and the temperature of the X-ray emitting
gas (e.g., Lubin \& Bahcall 1993)
we expect some correlation between \sn\ and cluster X-ray properties.
Figures~\ref{fig:snxprops} shows \sn\ plotted against 
cluster X-ray temperature $T_X$ (keV) and X-ray luminosity $L_X$
in the 0.5--4.5~keV band (ergs~sec$^{-1}$) from Jones \& Forman (1997).
While previous investigations found no correlations
of \sn\ with these properties (Harris \etal\ 1995; West \etal\ 1995),
our larger, more homogeneous data set clearly shows that there is
a correlation, with a high level of significance, $\sim\,$0.998.

The scatter in these plots is larger than in the velocity dispersion
plots, but that is at least partially due to the larger uncertainties
in the X-ray temperatures.  If the observed variation of \sn\ among these
galaxies is driven by environmental density, then \sn\
should correlate more strongly with $T_X$ than with $L_X$
because, assuming hydrostatic equilibrium, the gas temperature
is determined simply by the depth of the cluster gravitational potential
while the luminosity also depends on the amount of gas present.
These two plots look so similar because half of the temperatures
were estimated from the cluster $L_X$-$T_X$ relation (see
Table~\ref{tab:cls}), and there will be definite scatter about this
relation.  These estimated temperatures have $1\,\sigma$
uncertainties of about 20\%, while the uncertainties in
the velocity dispersions used above are $\lta\,$10\%.

\begin{figure} \epsscale{0.85}\plotone{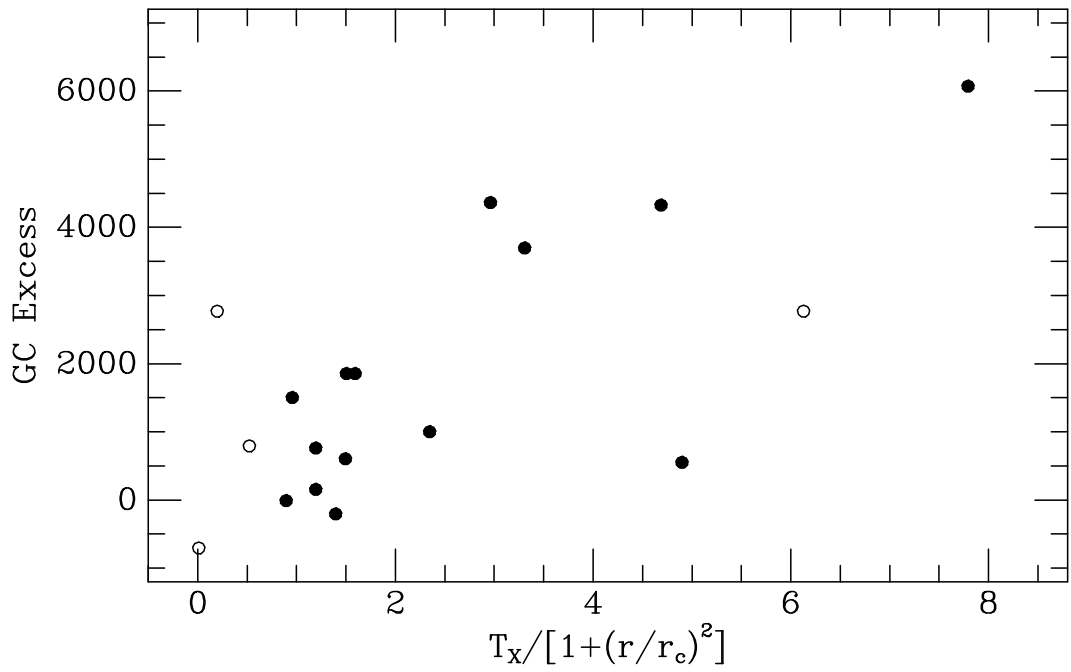}
\caption[fig13.ps]{ 
The observed ``excess'' number of GCs in the IGC model is plotted
against a quantity proportional to the projected matter density of the
cluster at the radial position of the galaxy. (See text for details.)
\label{fig:nexcess}}
\end{figure}

West \etal\ (1995) have revived the idea of
``intra\-cluster globular clusters'' (IGCs) which follow the overall
cluster mass profile rather than being bound to individual galaxies.
In this model, a giant galaxy which happens to lie near the cluster center
will appear to have a high \sn\ due to the ``excess'' intracluster
GCs which become {\it ipso facto} associated with it.
West \etal\ defined the excess to be the total number of GCs which elevate
\sn\ above a value of 4, and assumed that this will be proportional
to the projected matter density at the distance $r$ from the cluster
X-ray center. For hydrostatic equilibrium and an isothermal potential, 
the excess will be proportional to
$T_X/(1{\,+\,}r^2/r^2_c)$, where $r_c$ is the core radius of the cluster.
In Figure~\ref{fig:nexcess}, we plot the GC excess, defined relative to
$\sn{\,=\,}3.5$, which more accurately reflects the lower \sn\ limit for
our sample, against $T_X/(1{\,+\,}r^2/r^2_c)$, where the $r_c$ values come
from Jones \& Forman.  Consistent with the IGC model, we find
a roughly linear relationship.  
Even the non-central galaxies (open symbols) should follow this relationship, 
since it takes into account galaxy position in the cluster, and several 
of these galaxies do appear to deviate less than in the previous two plots,
although otherwise there is no improvement.
However, Blakeslee (1997a) showed that  $S_N$ correlated
marginally better with ``local X-ray luminosity'' $L_X/(1{\,+\,}r^2/r^2_c)$,
than with $L_X$ itself.

In the next section, we discuss
the implications of the observed correlations,
and non-correlations, for the various theoretical models.
We then describe how the correlations
found here improve our understanding of GC systems, and the mechanisms
which effect \sn\ in Abell cluster central galaxies.

\section{DISCUSSION}

\subsection{Are High-$S_N$ BCGs ``Special''?}

Among the bright cluster galaxies in this sample,
$S_N$ varies continously as a function
of certain well-defined galaxy and cluster parameters. 
Thus, the high-\sn\ galaxies do not constitute a special class
of object, but rather occupy the high end of the \sn\ distribution.
The total number of GCs appears to be determined primarily by cluster
environmental factors, unlike the central galaxy luminosity, which
is remarkably immune to variations in cluster properties (e.g., 
Hoessel, Gunn, \& Thuan 1980; PL).

As our results conflict with the common notion of ``normal'' and ``anomalous''
\sn\ systems, we give some consideration to the galaxies in the literature 
which have inspired it.  We begin with the prototypical high-\sn\ galaxy M87.
The most recent determination of \sn\ for this galaxy was
by McLaughlin \etal\ 1994, who reported a global value of $\sn=14.3\pm1.0$, 
before taking into account uncertainties in the distance or GCLF.
Those authors adopted the same distance as we have; however, their fitted
GCLF parameters significantly disagree with those determined with HST
(Whitmore \etal\ 1995).  This is understandable, as the fitted GCLF width
and turnover are strongly correlated when the data do not go significantly
fainter than $m^0$.  

The much larger 
field of the McLaughlin \etal\ study
makes their data preferable to HST for deriving \sn,
and they provided a correction factor, 
a function of the GCLF
parameters and distance modulus, that can be applied to their quoted \sn.
Plugging the HST GCLF results into their correction formula, we find
$\sn(M87){\,=\,}11\pm1$.
Moreover, \sn\ increases with radius in M87, so its ``metric \sn'', as we
report for our sample, should be smaller than this global value.
Thus, given the uncertainties, including the systematic error in
the calibration of Virgo to the CMB frame,
we can only conclude that M87 is at the high end of the continuous 
distribution in \sn\ which we observe.

The other well-studied galaxy that has been numbered among the high-\sn\
systems is the Fornax cD NGC~1399.  This galaxy was first reported to have 
$\sn{\,\sim\,}16$ (Bridges, Hanes, \& Harris 1991).  A more recent study,
using GCLF parameter which agree with ours (Blakeslee \& Tonry 1996)
found $\sn{\,=\,}12\pm3$ (Kissler-Patig \etal\ 1996).  However, this
later study used a distance modulus that was
0.30~mag smaller than the new HST Cepheid result for this cluster
(Silberman \etal\ 1996).
Using the Cepheid distance modulus, \sn\ for this galaxy becomes
\sn(N1399)${\,=\,}9\pm2$. The GCs and halo light follow similar distributions,
so the metric \sn\ would not differ much from the global value.  This
lies in the range of the galaxies studied here, though higher than
we would expect for a cluster of low dispersion.
On the other hand, Fornax is not a rich cluster (showing up as S373 in 
the supplementary catalog of southern poor clusters of
Abell \etal\ [1989]), though it is very compact,
 so it is not clear how the \sn\ of its central
galaxy should compare to those studied here.

The rest of the high \sn\ values found in the literature
generally scale with the \sn\ of M87.  For instance, 
Harris \etal\ (1995) find 22$\pm$7 and 13$\pm$6 for the central
cD galaxies in A2052 and A2107 (both outside the redshift limit of
our survey; their other cluster, A2666, is discussed below).
These numbers were derived relative to an assumed M87 value of 15;
using the value of 11 implied by the HST GCLF yields
\sn(A2052)${\,=\,}16\pm5$ and \sn(A2107)${\,=\,}9.5\pm4.4$, which are both
still high, but not even the A2052 \sn\ can be called anomalous, given
the uncertainties.

Direct comparisons between our results and literature values
are possible for several galaxies. 
\cite{bt95} used RC2 photometry and reported global values of \sn\ which 
were in agreement with those found by Harris (1987), but
we have revised those numbers down based on better 
photometry.  In fact, Harris also used RC2 photometry, so his \sn\ numbers
get revised down in an identical manner; in particular, his value for
NGC 4874 becomes $9\pm4.5$ when RC3 photometry is used.
Butterworth \& Harris studied A1367-1 and reported the ratio 
\sn(A1367)/\sn(M87) as a function of radius.
Reading the value of this ratio at our metric radius of
$\sim\,$1\farcm8, and using the M87 \sn\ from above gives
$\sn=5.2\pm1$, very close to our value; even if we had
used the larger \sn\ of 14 for M87, we would agree with their
results within the errors.
For A2666-1, Harris \etal\ (1995) found  $\sn=3\pm2$, again assuming the
\sn\ of 15 for M87.  Using the smaller M87 value, their A2666-1 \sn\
drops to $\sim\,$2.2.  In either case, it is close to our value
of 3.5 for this galaxy.
Finally, for the A2199 cD NGC~6166, Pritchet \& Harris (1990) reported
$S_N{\,\lta\,}4$, but a redetermination based on improved data
found $\sn{\,=\;}9^{+9}_{-4}$ (Bridges \etal\ 1996).
The latter result agrees with our value of \sn${\,=\,}$8.1 for this galaxy.

In summary, for the few cases in which direct comparisons are possible,
our results are in good agreement with published values.
However, we believe that our larger, more homogenous data set affords
a clearer view of the overall picture.  
We find that there is a continuum of possible \sn\ 
values for bright cluster galaxies.
Where a galaxy falls in this continuum may even be predictable
from the cluster velocity dispersion, or $T_X$, and the projected
distance of the galaxy from the cluster center.

\subsection{How Do the Models Compare?}

The observations presented here provide an unprecedented amount of
information about \sn\ variation among central galaxies
in rich clusters; thus,
we have the opportunity to evaluate the theoretical models from 
a more favorable vantage point than previously accessible.
The history of GC formation theories is a long one, dating back
at least to the work of Peebles \& Dicke (1968), but we will consider
in turn only those extant models which purport to explain the
wide variation of \sn\ among central galaxies in clusters. 
A good model may even guide our understanding of the observations.

(1) Initial conditions.~~Until now, it was thought that
no obvious correlations existed between the \sn\ of the central
cluster galaxy and other properties of either the galaxy or cluster
(West \etal\ 1995).  For instance, Harris et al.\ (1995), found no
correlation of \sn\ with
cluster X-ray properties in a data set composed of their three 
galaxies and others from the literature, including central galaxies
in 4 more Abell clusters, 3 AWM and MKW poor clusters, Fornax, and 
Virgo.  They interpreted their results as indicating that
GC formation in central cluster galaxies occurred early on, and reflected
local initial conditions, thus the lack of any correlations.
This view, that modern galaxies with high values of \sn\ ``were special
{\it ab initio}'' (van den Bergh 1984) and consequently formed
their GCs ``superefficiently'' (McLaughlin \etal\ 1993),
we call the ``initial conditions'' scenario. 
It goes back to Harris (1981) and is the prevailing view in the literature
(e.g., Harris 1991).

In this scenario one must explain why some protogalaxies
were privileged to undergo GC formation with enhanced efficiency.
Harris \& Pudritz (1994) constructed a model
for GC formation out of primordial pressure-confined, self-gravitating,
magnetized supergiant molecular clouds.  They hypothesized that 
larger external pressures may have prevailed in the proto-halos of
cD galaxies and caused the primordial clouds in their
model to fragment into more proto-GC cores per unit cloud mass.
This idea remains speculative. In any case, our data indicate that many of
the reasons for invoking initial conditions are no longer
valid.  In particular, it is {\it not} true that there are no correlations
of \sn\ with cluster properties.  We find very
good correlations between \sn\ and cluster X-ray properties and with
cluster velocity dispersion.   Thus, we do not believe that the
central galaxies in our study formed their GCs ``superefficiently'' 
as a result of local initial conditions.

(2) Biasing.~~A similarly motivated, but fundamentally different model is 
that of ``biased GC formation.''  This is the apparent heir of the early
work by Peebles \& Dicke, the major addition being that of dark matter.
Examples of this include the models of Peebles (1984) and West (1993).
In the biasing scenario, GC formation depends only on the height of the
local density fluctuations with respect to some universal threshold.
When the small scale (pc-sized) fluctuations are superposed on a larger
scale (Mpc-sized), low amplitude fluctuation, such as might eventually
evolve into a cluster of galaxies, the small scale fluctuations are more
likely to broach this universal threshold; thus, they have a ``bias''
working in their favor. Biased formation may even produce GCs 
outside of any particular protogalaxy, with the resulting intergalactic
GCs becoming associated with the whole of the cluster that forms there.
Thus, the biasing model is also primordial, but less ``local''
than the initial conditions model.
It predicts that \sn\ should correlate strongly with the 
present-day density of the local environment, as we observe,
but we are biased against this model because it is largely unconstrained.
Biasing has been proposed as a possible mechanism at work in the
IGC model discussed below; we will come back to it.

(3) Mergers.~~In this model, gas-rich galaxies with low \sn\ values, spirals
in particular, merge to form elliptical galaxies, as suggested by
Toomre (1977). In the process, GCs form out of shocked gas clouds,
increasing the value of \sn\ with every merger by an amount
proportional to the available gas (Ashman \& Zepf 1992; Schweizer 1987).
Repeated mergers of such systems in the core of a galaxy cluster 
might then yield high values of \sn.
Kumai \etal\ (1993a,b) have discussed mechanisms for creating GCs 
in gas-rich collisions.

While mergers do indeed occur and massive
star clusters may form in the process (e.g., Holtzman 1992),
the question we must address is whether or not the observed
variations in \sn\ among central galaxies in clusters can be explained
in this way.  As formulated (Ashman \& Zepf 1992; Zepf \& Ashman 1993),
the merger model predicts a strong dependence
of \sn\ on galaxy luminosity, which we do not see,
while we do see other correlations which are not obvious
in this picture.  Thus, we do not believe that mergers of modern day
galaxies can explain the observations.  However, mergers of gaseous
fragments in the protogalactic era, consistent with hierarchical
structure formation models, almost certainly played a role in determining
\sn, as we discuss in \S\ref{sec:mymodel}.

(4) Stripping.~~The idea that the central galaxy in the cluster potential 
increases its \sn\ through preferential tidal stripping of the GCs from
other galaxies in the cluster goes back to Forte, Martinez, \& Muzzio (1982).
Muzzio (1987) reviewed the early work done on modeling such GCS
dynamical evolution.
Since the GC populations of elliptical galaxies tend to be more extended than
the halo light (e.g. Harris 1991), a central galaxy might increase its
\sn\ through this process; \cite{bt95} noted that the number of GCs
donated by other galaxies would not be unreasonable.
The observed strong correlation of central galaxy \sn\ with cluster
velocity dispersion might occur as a result of stripping, since the 
crossing time in high dispersion clusters is lower and each 
galaxy would have passed through the core more times (although 
higher velocities also make stripping less efficient).  Moreover,
the correlations with local galaxy density and location of the galaxy
relative to the cluster center are both easily understandable 
in the context of this model.  One problem is that
the increase of \sn\ for the central cluster galaxy was too slow
in the simulations reviewed by Muzzio (1987) to explain the observations.
More modern simulations would be desirable.
Stripping in the context of the IGC model is discussed below.

(5) Intra\-cluster globular clusters (IGCs).~~The IGC
model of West \etal\ (1995), mentioned above
when we examined correlations of \sn\ with cluster properties,
simply proposes that there are large populations of GCs that belong to
the cluster as a whole and follow the overall cluster density in their
distribution.  West \etal\ suggested stripping and biasing as possibilities
for the origin of the IGCs.

As stated previously, our results appear consistent with the prediction
of the IGC model that the number of GCs in excess of \sn$\,\sim\,$4 should
correlate with $T_X/(1{\,+\,}r^2/r^2_c)$.  West \etal\
predicted that the \sn\ values of two of our sample galaxies, A569-1
and A779-1, would be relatively low, \sn$\,\sim\,$4, as we find in both cases.
We believe that the IGC model is on the right track in treating
the GC populations of the central galaxies as more a property of
the cluster itself than of the galaxy.  However, in order to clump
around the galaxy, the IGCs must have a velocity dispersion
closer to the internal dispersion of the galaxy ($\sim\,$300 \kms) than of
the cluster ($\sim\,$750 \kms).  This is observed to be the case for M87 where
the dispersion of the GCs is $\sim\,$400 \kms\ (Cohen \& Ryzhov 1997).
Moreover, the origin of the IGCs remains problematic.  Biasing
suffers the problems endemic to dark matter models in general, namely the lack
of any useful external constraints. The alternative mechanism of stripping
qualitatively explains our observations, but may not
be efficient enough to produce large IGC populations.
Therefore, it is not clear that the GCs are truly intergalactic;
what we mean by saying that they are ``more a property of the cluster''
is that their number is determined by cluster properties, not by the
properties of the central~galaxy.

We conclude that no clear winner has emerged from among the available
models, although there have been some strong contenders.  In the following
section we use as an additional guide the observationally known properties
of the BCGs themselves.  Along with the insight
gained from our observations and this discussion, i.e., that
the central galaxy GCs are more rightly considered a cluster property,
we attempt to reach a coherent understanding of
the observed variations in \sn.

\subsection{Another Model}\label{sec:mymodel}
\def\ml{{\it M$/$L}}

Discussions about why only certain central galaxies in clusters
have ``anomalously'' high values of \sn\ usually revolve around
the presence of a cD envelope, which appears to be a helpful but
not sufficient condition for having a high \sn.  In an effort to 
uncover further clues in this regard, we extend the
discussion to include another remarkable property of central cluster
galaxies, namely their uniformity.

Sandage (1972) was the first to exploit the small dispersion in 
the metric luminosities $L_m$ of brightest cluster galaxies.
He reported that $L_m$ defined within $\sim\,$20 \hkpc\ had a dispersion
of only 0.25~mag for BCGs and was independent of cluster richness. 
Hoessel, Gunn, \& Thuan (1980) studied a larger, unbiased sample of 
BCGs and found an intrinsic dispersion of \hbox{$\sim\,$0.35} mag
for $L_m$ defined within a 10 \hkpc.  They also found that $L_m$ did
become slightly brighter on average with cluster richness.
Hoessel (1980) showed that this trend 
also correlated with the structure parameter $\alpha$, the slope of the
luminosity profile at the metric radius. He defined the $L_m$-$\alpha$ 
distance indicator, which had no residual dependence on richness.
As Oegerle \& Hoessel (1991) showed, this indicator gives distances
for BCGs that are nearly as good as 
fundamental plane distances but much less expensive observationally.

Most recently, Lauer \& Postman (LP 1994; PL 1995) have reinvestigated the 
use of the BCG $L_m$-$\alpha$ distance indicator for a sample of 119 BCGs,
from which our own sample was selected. 
PL find that the intrinsic dispersion in $L_m$
drops from 0.33 mag to 0.24 mag after the $\alpha$ correction has been
applied, and residuals about the $L_m$-$\alpha$ relation show no dependence
on BCG luminosity, color, or location, nor on cluster richness.
Moreover, they report that $L_m$ is independent of cluster richness even
before the $\alpha$ correction has been applied. 
(Although Hudson \& Ebeling [1997] find evidence that both $L_m$ and residuals
from the $L_m$-$\alpha$ indicator correlate with cluster X-ray luminosity.)
They conclude that
the small scatter in $L_m$ and $(B{-}R)$ color, coupled with the lack of any
second-parameter effects, make BCGs ``the most homogeneous distance
indicators presently available for large-scale structure research.''

In contrast to the uniformity of
BCG metric luminosity, the \sn\ values of BCGs have a range
of more than a factor of three and are strongly correlated
with cluster density as measured by vel\-ocity dispersion, X-ray temperature,
and local galaxy density.
Taken together, these points imply that high-\sn\ BCGs
are not anomalously rich in GCs,
rather, they are underluminous with regard to their preeminent positions
at the centers of rich clusters, while the numbers of GCs 
accurately reflect the dense environments.

\begin{figure}\epsscale{0.85}\plotone{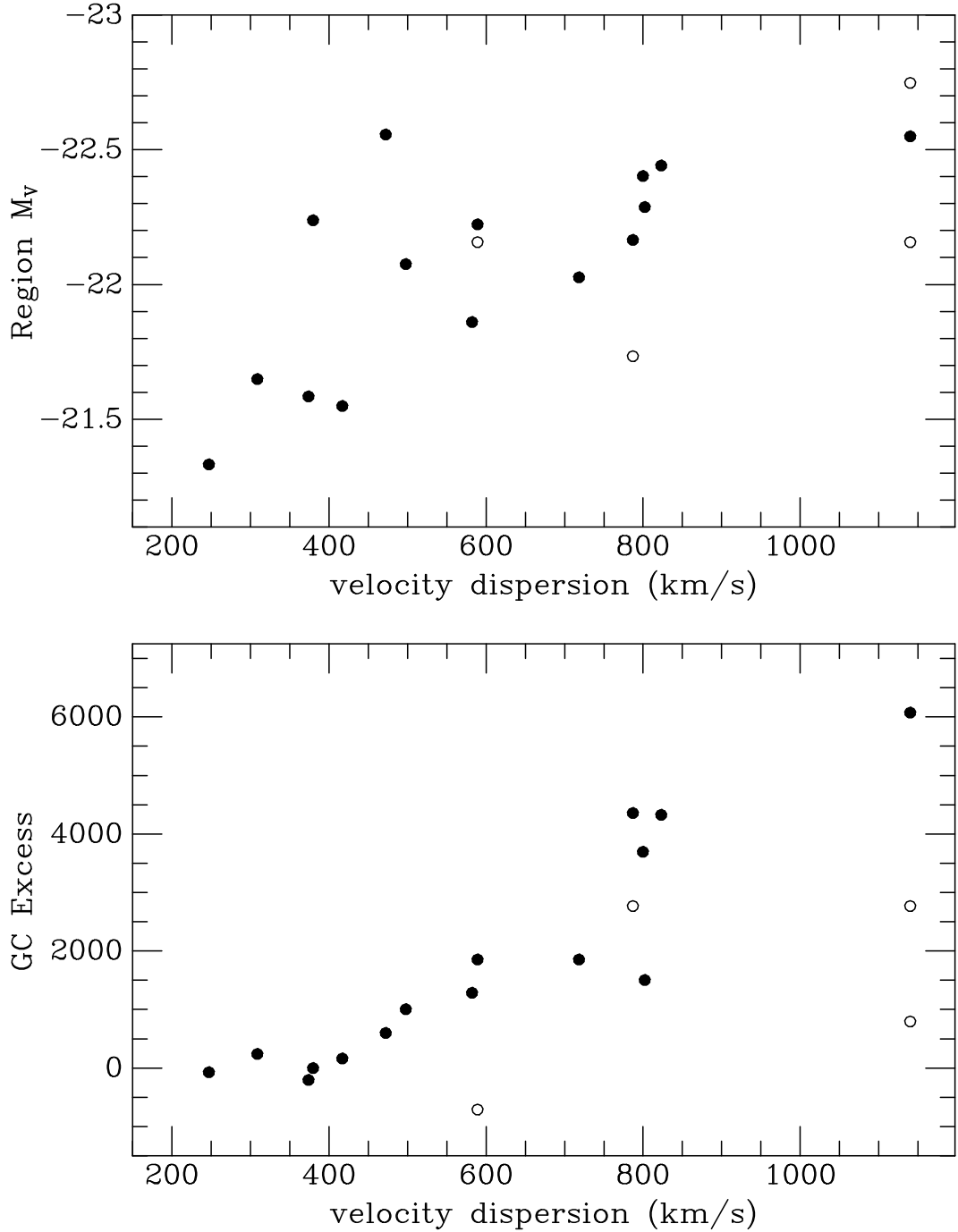}
\caption[fig14.ps]{ 
A comparison of the sensitivities of galaxy luminosity and GC number
to cluster density.
In $(a)$, the total absolute $V$ magnitude of the region in each galaxy 
over which
our metric $S_N$ values have been meausured (see Figure~\ref{fig:sn_mag})
is plotted against cluster velocity dispersion. 
In $(b)$, the excess GC number above $\sn{\,=\,}3.5$ is plotted
against cluster velocity dispersion.  
\label{fig:discuss}}
\end{figure}

To illustrate this view, Figure~\ref{fig:discuss}$a$ plots absolute magnitudes
within 40 kpc (excluding galaxy center, 
as in the lower panel of Figure~\ref{fig:sn_mag})
against cluster velocity dispersion.  There is a general trend, but
the intrinsic scatter is large (the uncertainty in $M_V$ is $\sim\,$0.1~mag).
In Figure~\ref{fig:discuss}$b$, we show ``GC excess'' (the number of
GCs which elevate \sn\ above a value of 3.5) plotted against cluster
dispersion.    Despite the larger observational
uncertainty, the scatter is much smaller, and
the only discrepant points are non-central galaxies.
Dividing the smoothly varying GC excesses in Figure~\ref{fig:discuss}$b$ by 
the relatively invariant luminosities in Figure~\ref{fig:discuss}$a$ 
(and adding 3.5)
results in the observed correlations of \sn\ with cluster density.
The correlation of \sn\ with $\alpha$ (Figure~\ref{fig:sn_ext}) can
then be seen as a consequence of the fact that $\alpha$ itself
correlates with cluster density (Blakeslee 1997a).

These observations might be explained by a 
model in which the GCs of these galaxies, like those of the Milky Way,
formed early on and in proportion to the available mass.  
A later process, perhaps the collapse of the surrounding cluster,
then halted the luminosity growth of the central galaxy.
If this is true, then the number of GCs per unit total ``background''
(cluster) mass interior to a projected radius $R$ provides a more accurate
description of the GC formation process than does the number per
unit luminosity (i.e., \sn).  We will attempt to estimate the cluster
mass within 40 kpc, and so derive the number of GCs per unit mass,
a quantity we call $\eta_{_{GC}}$, for the galaxies in our sample.

\begin{figure} \epsscale{0.9}\plotone{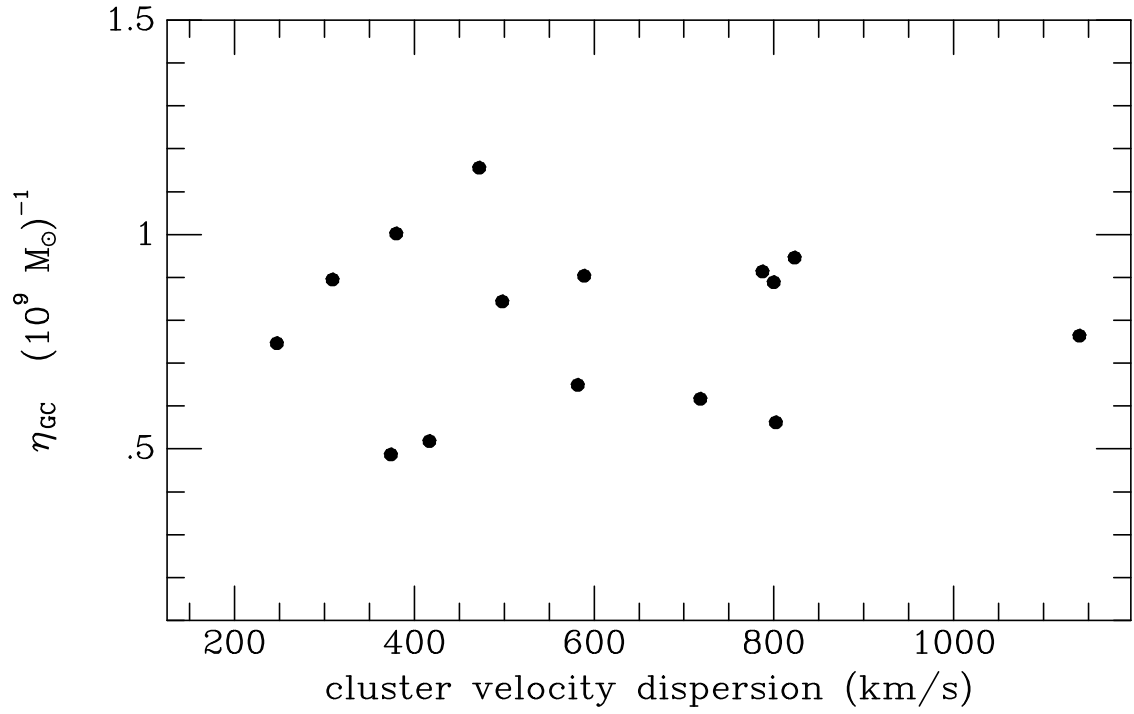}
\caption[fig15.ps]{ 
The number of GCs per unit 10$^9\,M_{\odot}$,
denoted $\eta_{_{GC}}$, is plotted against cluster
velocity dispersion for the central cluster galaxies in our sample.
To estimate the total mass interior to $R{\,=\,}40$~kpc, we used 
a cluster core model normalized to the central density
of a non-singular isothermal sphere.
$\eta_{_{GC}}$ calculated in this way is constant.\label{fig:Ngc_disp}}
\end{figure}

The surface mass density near the center of a nonsingular isothermal
sphere is 
$\Sigma_c = 9 v^2_{rms}/2 G r_c$, 
where $v_{rms}$ is the cluster velocity dispersion,
and $r_c$ is the core radius.
If $r_c$ were the same for all clusters, or were completely uncorrelated
with $v_{rms}$, then the mass enclosed within a fixed radius would go
simply as the square of the velocity dispersion.  However, there is 
a rough relationship between $v_{rms}$ and $r_c$, in the sense
that clusters with larger core radii tend to have higher velocity dispersions.
Because of the difficulty in accurately measuring $r_c$ and the fact
that a quarter of our clusters with known dispersions lack $r_c$ values,
we will make use of this relationship.
A straight fit for our sample gives $r_c \approx 2.5\,v_{rms}^{0.6}$,
where $r_c$ is in kpc, $v_{rms}$ is in \kms, and the scatter is about 50\%.
Using this relation, we get the following expression for the total mass
enclosed within 40~kpc of the central galaxy
\begin{equation}\label{eq:isosig}
M_{c}(R{\,=\,}40{\,\rm kpc}) \, \approx \, 4.0 \times 10^{12} \, 
  {\left( v_{rms} \over 500\,\hbox{km~s$^{-1}$}\right)}^{1.4} M_{\odot} \, .
\end{equation}
Dividing the observed number of GCs within 40 kpc by the mass implied
by Eq.~(\ref{eq:isosig}) yields the ratio $\eta_{_{GC}}{\,\equiv\,}N_{GC}/M_c$,
which we plot against cluster velocity dispersion in Figure~\ref{fig:Ngc_disp}.

The figure demonstrates that $\eta_{_{GC}}$ calculated according to
this simple mass model is constant.
The average value is about 0.7~per 10$^9\,M_{\odot}$,
with a scatter of $\sim\,$30\%.
Thus, there appears to have been a single, universal formation efficiency
for the GCs in our sample.  For 10$^6\,M_{\odot}$ GCs, the implied
efficiency would be $\lesssim\,$0.1\%.
Another way to describe the situation is to note that
the observed $N_{GC}$ goes approximately as $v_{rms}^{1.4}$. 
Because we wish to avoid models which postulate dark matter
biasing or a density dependent efficiency, 
we hypothesize that the production of GCs per unit mass is constant.
Finally, we have shown that a simple,
plausible mass model produces the desired result.

\section{SUMMARY}

We have studied the largest sample to date of
GC populations around central galaxies in Abell clusters, 7~times
larger than any previous single study.  
The primary factor which allowed us to study such a large sample
with a relatively small telescope was our use of the analysis methods
developed by \cite{bt95}.
Our sample was selected from the BCG survey of Lauer \& Postman (1994);
nearly all the new observations were made
in the $R$-band, the one exception being that of A2199. 
For several clusters, the second brightest galaxy 
was included because it was comparable to the BCG in luminosity.
In fact, in all of these cases, the second brightest galaxy
was also more centrally located, as judged
by both position relative to the X-ray center and the local galaxy
density.  In addition, a third Coma galaxy, NGC 4839 (A1656-3), was included
by virtue of its cD envelope.

We detected a GC system and presented a measurement of the GCLF width
$\sigma$ for each sample galaxy.
A slight bias was found in the poorly constrained values of $\sigma$,
but the more tightly constrained values
showed little intrinsic dispersion and agreed closely with
those found from deep imaging of nearby ellipticals, particularly
M87, which was used as our ``\mo\ calibrator.'' 
We noted that our technique, as well as previous studies which
used only direct counts, relied to some degree on the predictability
of the GCLF turnover magnitude \mo.  That is, the frequently made,
but poorly tested, assumption of a universal GCLF is an important input.
Thus, our GCLF results confirm the
working assumption of a universal GCLF for central galaxies in
the cores of rich clusters.

To limit biased comparisons of \sn\ among sample galaxies, we assumed
a single value of $\sigma$, consistent with our data, and calculated
metric values of \sn, defined within 40~kpc.  Our results for \sn\
followed a continuous distribution; they were not segregated into ``normal''
and ``high'', or ``anomalous'', classes.   This was the first time in
which a continuum of \sn\ values was clearly shown to exist.
Contrary to the results of other studies, we find that \sn\
correlates well with properties of the galaxy clusters, particularly
central and asymptotic velocity dispersions, but also X-ray luminosity and
temperature and the local galaxy density.  For clusters with multiple
galaxies in our sample, the one with the higher \sn\ value
was always the one which was more extended and closer to
the cluster X-ray center.  We did not find a strong correlation
with cluster richness expressed in Abell counts, but that is not
surprising, given their fairly qualitative and uncertain nature.
We found no evidence of a previously proposed correlation of
central galaxy \sn\ with Bautz-Morgan class, and there is
no obvious correlation with Rood-Sastry type either.

We discussed several models which have been proposed to explain
the \sn\  values of central cluster galaxies.
Our data confirmed some predictions of the ``intracluster globular
cluster'' (IGC) model, but the lack of a viable mechanism
for producing the IGCs was judged problematic.  The stripping
model appeared most consistent with our data, but it may require
an unacceptably long time scale for adding GCs to the central galaxy.

We concluded by offering our own explanation for the behavior of
\sn\ in these galaxies. 
We suggested that the GCs formed early, with their number
$N_{GC}$ scaling in proportion to the available mass.  
(We do not exclude the possibility that their formation may
have been episodic, yielding the frequently observed
complex color/metallicity distributions.)
The galaxy luminosity, on the other hand, is relatively independent
of the cluster mass; perhaps the BCG formation was punctuated by the
growth of the surrounding cluster in such a way as to
produce the observed ``standard candle'' quality of these galaxies.
In any case, the result is that \sn,
the ratio of $N_{GC}$ to galaxy luminosity,
is observed to scale with cluster mass in the same way as $N_{GC}$ 
itself does. 

A better diagnostic of the physical processes affecting GC formation
should therefore be provided by the observed number of GCs per unit 
``background'' mass, a quantity which we have called $\eta_{_{GC\,}}$.
To estimate it, we adopted a simple flat core model and used 
it to calculate the total mass within the volume occupied by the GCs.
We showed that the resulting $\eta_{_{GC\,}}$ for our sample of galaxies
was constant in this model.  The implied ``universal'' formation
efficiency is $\lesssim\,$1~GC per 10$^9 M_\odot$.   
Thus, we believe that a consistent picture of the GC systems of 
central galaxies in clusters has begun to emerge, although further
observational and theoretical work is needed 
in order to verify these results.

\acknowledgments

We thank Bob Barr for keeping MDM Observatory
running smoothly and Paul Schechter for many helpful comments.
We are grateful to Mark Postman and Tod Lauer for electronic data
on their BCG sample, Guy Worthey for redshifted stellar populations
calculations, Christine Jones-Forman for cluster X-ray information,
and Jeff Willick for double-checking the Coma-Virgo relative distance
from the Mark~III Catalog.
We thank Bjarne Thomsen, the referee, for numerous comments
and suggestions that improved the paper.
This research was supported by NSF grant AST94-01519, and the paper
was completed under the sponsorship of a Caltech Fairchild Fellowship.

\clearpage

\begin{deluxetable}{rccrcccl}
\small\centering\tablewidth{0pt}
\tablecaption{The BCG Sample\label{tab:bcg}}
\tablehead{
\colhead{Abell} & \colhead{RA~(J2000)~Dec}& \colhead{~$l$~~~~~~~~~$b$} & \colhead{$cz_{h}$} & 
\colhead{$A_B$}& \colhead{~$M_{R,m}$} &\colhead{$\alpha$} & \colhead{Name~~~~~~~~} }
\startdata
  262-1&  01 52 46~~$+$36 09 05 & 136.57~$-$25.09   & 4831 & .24 &$-$22.189 & .810 & N0708  \\
  347-1&  02 25 26~~$+$41 49 27 & 141.11~$-$17.71   & 5257 & .24 &$-$22.352 & .601 & N0910  \\
  397-1&  02 56 29~~$+$15 54 59 & 161.81~$-$37.34   &10286 & .27 &$-$22.542 & .582 & C463-037 \\ 
  539-1&  05 16 55~~$+$06 33 10 & 195.65~$-$17.60   & 9682 & .51 &$-$22.484 & .511 & C421-019 \\ 
  539-2&  05 16 37~~$+$06 26 28 & 195.71~$-$17.72   & 8318 & .51 &$-$22.210 & .785 & U03274\\
  569-1&  07 09 08~~$+$48 36 55 & 168.58~$+$22.80   & 5724 & .34 &$-$22.418 & .486 & N2329 \\
  634-1&  08 15 45~~$+$58 19 16 & 159.06~$+$33.79   & 8135 & .13 &$-$22.258 & .498 & U04289\\
  779-1&  09 19 47~~$+$33 44 59 & 191.09~$+$44.39   & 6867 & .00 &$-$22.858 & .594 & N2832\tablenotemark{\dagger} \\ 
  999-1&  10 23 24~~$+$12 50 06 & 227.94~$+$52.58   & 9749 & .11 &$-$22.267 & .441 & C065-015 \\ 
 1016-1&  10 27 08~~$+$11 00 37 & 231.29~$+$52.48   & 9705 & .03 &$-$22.048 & .430 & I0613  \\
 1177-1&  11 09 44~~$+$21 45 32 & 220.44~$+$66.29   & 9561 & .00 &$-$22.453 & .724 & U06203\\
 1185-1&  11 10 38~~$+$28 46 03 & 202.81~$+$67.72   &10521 & .00 &$-$22.444 & .616 & N3550 \\
 1314-1&  11 34 49~~$+$49 04 38 & 151.77~$+$63.54   & 9977 & .00 &$-$22.461 & .583 & I0712  \\
 1367-1&  11 44 02~~$+$19 56 59 & 234.29~$+$72.99   & 6237 & .00 &$-$22.496 & .518 & N3842 \\
 1656-1&  13 00 08~~$+$27 58 36 &\phn57.19~$+$87.89 & 6497 & .05 &$-$22.957 & .590 & N4889 \\
 1656-2&  12 59 36~~$+$27 57 33 &\phn58.08~$+$88.01 & 7176 & .05 &$-$22.545 & .855 & N4874\tablenotemark{\dagger} \\ 
 1656-3&  12 57 25~~$+$27 29 48 &\phn48.79~$+$88.62 & 7335 & .03 &$-$22.287 & .610 & N4839\tablenotemark{\dagger} \\ 
 2162-1&  16 12 36~~$+$29 29 04 &\phn48.33~$+$46.01 & 9547 & .07 &$-$22.475 & .503 & N6086\tablenotemark{\dagger} \\ 
 2197-1&  16 29 45~~$+$40 48 42 &\phn64.68~$+$43.51 & 8800 & .00 &$-$22.887 & .586 & N6173 \\
 2197-2&  16 27 41~~$+$40 55 37 &\phn64.84~$+$43.90 & 9408 & .01 &$-$22.350 & .702 & N6160\tablenotemark{*} \\
 2199-1&  16 28 38~~$+$39 33 03 &\phn62.93~$+$43.69 & 9348 & .00 &$-$22.657 & .755 & N6166\tablenotemark{\dagger} \\ 
 2634-1&  23 38 29~~$+$27 01 50 & 103.50~$-$33.07   & 9141 & .16 &$-$22.748 & .650 & N7720\tablenotemark{\dagger} \\ 
 2666-1&  23 50 59~~$+$27 08 48 & 106.72~$-$33.81   & 8123 & .13 &$-$22.768 & .549 & N7768 \\
\enddata
\tablenotetext{\dagger}{Schombert (1988) cD galaxy.}
\tablenotetext{*}{Selected as BCG by Hoessel, Gunn, \& Thuan (1980).}
\end{deluxetable}

\begin{deluxetable}{rccrrccccc}
\small
\centering
\tablewidth{0pt}
\tablecaption{Abell Cluster Information\label{tab:cls}}
\newdimen\digitwidth
\newdimen\minuswidth
\setbox0=\hbox{\rm0}
\digitwidth=\wd0
\catcode`?=\active
\def?{\kern\digitwidth}
\setbox0=\hbox{\rm-}
\minuswidth=\wd0
\catcode`!=\active
\def!{\kern\minuswidth}
\tablehead{
\colhead{Abell}& \colhead{?RA$_X$ (J2000) Dec$_X$} & \colhead{$cz_h$} & \colhead{$cz_{C}$} & 
\colhead{$cz_{A}$} & \colhead{?$\sigma$} &\colhead{$T_X$} & \colhead{R} & \colhead{BM} &\colhead{RS}}
\startdata
  262 &	?1 52 46~~+36 08 36 & 4913  &  4659  &  5310  &  ?498 & 2.4    & 0 &  III    & C  \\
  347 &	??\dots ??~~????\dots ?? & 5604 & 5391 & 6000  & ?582 & ... & 0 &  II-III    & C  \\
  397 &	?2 56 38~~+15 53 38 & 9975  &  9765  & 10560  &  ?... & 1.6\tablenotemark{*} & 0 &  III    & F  \\
  539 &	?5 16 36~~+06 26 30 & 8754  &  8755  &  9390  &  ?787 & 3.0    & 1 &  III    & F  \\
  569 &	?7 09 11~~+48 36 58 & 5749  &  5832  &  6060  &  ?374 & 1.4\tablenotemark{*} & 0 &  II     & B  \\
  634 &	?8 14 34~~+58 02 52 & 8135  &  8234  &  8280  &  ?309 & 0.9\tablenotemark{\dagger} & 0 &  III    & F  \\
  779 &	?9 19 47~~+33 44 49 & 6796  &  7039  &  6930  &  ?472 & 1.5\tablenotemark{*} & 0 & I-II    &cD  \\
  999 &	10 23 23~~+12 50 13 & 9603  &  9942  &  9600  &  ?417 & 1.2\tablenotemark{*} & 0 &II-III   & L  \\
 1016 &	10 27 03~~+10 58 42 & 9669  & 10013  &  9660  &  ?247 & 1.3\tablenotemark{\dagger} & 0 &  ...    & L  \\
 1177 &	??\dots ??~~????\dots ?? &  9561  & 9885  & 9420 &  ?... & ... & 0 &  I      &cD  \\
 1185 &	11 10 45~~+28 42 46 & 9917  & 10217  &  9780  &  ?718 & 3.9    & 1 &  II     & C  \\
 1314 &	11 34 48~~+49 05 10 & 9838  & 10043  &  9690  &  ?... & 5.0    & 0 &  III    & C  \\
 1367 &	11 44 40~~+19 42 35 & 6469  &  6795  &  6240  &  ?802 & 3.5    & 2 &II-III   & F  \\
 1656 &	12 59 43~~+27 56 12 & 6961  &  7229  &  6570  &  1140 & 8.1    & 2 &  II     & B  \\
 2162 &	??\dots ??~~????\dots ?? & 9629 &  9689  & 9030 &  ?... & ...  & 0 &  II-III & I  \\
 2197 &	16 27 40~~+40 55 39 & 9042  &  9065  &  8520  &  ?589 & 1.6\tablenotemark{*} & 1 &  III    & L  \\
 2199 &	16 28 38~~+39 33 10 & 9034  &  9059  &  8490  &  ?823 & 4.7    & 2 &  I      &cD  \\
 2634 &	23 38 25~~+27 00 56 & 9153  &  8807  &  9330  &  ?800 & 3.4    & 1 &  II     &cD  \\
 2666 &	23 51 01~~+27 08 25 & 8057  &  7714  &  8250  &  ?380 & 0.9\tablenotemark{*} & 0 &  I      &cD  \\
\enddata
\tablenotetext{*}{Temperature estimated from X-ray luminosity.}
\tablenotetext{\dagger}{Upper limit.}
\end{deluxetable}

\begin{deluxetable}{cccccccl}
\small\centering\tablewidth{0pt}
\tablecaption{Observing Runs\label{tab:runs}}
\tablehead{
\colhead{Run} & \colhead{Detector} & \colhead{$^{\prime\prime}/pix$} &\colhead{F}& 
\colhead{$m_1$} & \colhead{$A$} & \colhead{$C$} & \colhead{Notes} }
\startdata
0593 & Loral 2048$^2$& 0.343 & $I$ & 24.349 & 0.140 & 0.030 & Coma; 2$\times$2 \\
0794 & STIS 2048$^2$ & 0.240 & $I$ & 24.034 & 0.085 & 0.020 & N6166\\
1194 & Tek 1024$^2$  & 0.275 & $R$ & 25.582 & 0.098 & 0.011 & poor weather \\
0295 & Tek 1024$^2$  & 0.275 & $R$ & 25.565 & 0.111 & 0.015 & \\
0395 & Tek 1024$^2$  & 0.275 & $R$ & 25.530 & 0.115 & 0.014 & bright moon \\
0995 & Tek 1024$^2$  & 0.275 & $R$ & 25.340 & 0.117 & 0.015 & \\
\enddata
\end{deluxetable}

\begin{deluxetable}{rcrccccc}
\small\centering\tablewidth{0pt}
\tablecaption{Galaxy Observations\label{tab:obs}}
\tablehead{
\colhead{Galaxy} & \colhead{Run} & \colhead{Exp} & \colhead{\pf} & 
\colhead{\mstar} & \colhead{$\mu_{sky}$} & \colhead{\snrz} & 
\colhead{N$_{\rm obj}$} }
\startdata
 A262-1 & 1194 &  5600 & 1.10 & 34.681 & 20.75 &  4.4  & 1394 \\
 A347-1 & 1194 &  4000 & 0.99 & 34.348 & 20.67 &  3.0  & 1320 \\
 A347-1 & 0995 &  4200 & 0.86 & 34.145 & 21.05 &  3.8  & 2351 \\
 A397-1 & 0995 & 20700 & 0.87 & 35.852 & 20.29 &  1.8  & 1350 \\
 A539-1 & 0295 & 11900 & 0.93 & 35.348 & 20.92 &  2.2  & 1300 \\
 A539-2 & 0295 & 10850 & 0.89 & 35.250 & 20.77 &  2.0  & 1511 \\
 A569-1 & 0295 &  4200 & 0.96 & 34.335 & 21.18 &  3.3  & 1210 \\
 A569-1 & 0395 &  6000 & 1.24 & 34.656 & 20.96 &  2.7  &\phn867 \\
 A634-1 & 0295 & 12600 & 0.98 & 35.626 & 21.24 &  3.1  & 1796 \\
 A779-1 & 0295 &  6600 & 0.90 & 35.018 & 21.14 &  3.3  & 2086 \\
 A999-1 & 0295 & 13200 & 0.88 & 35.679 & 21.12 &  2.3  & 1802 \\
A1016-1 & 0295 & 11700 & 0.86 & 35.597 & 21.09 &  2.2  & 1590 \\
A1177-1 & 0295 & 15025 & 0.96 & 35.903 & 21.14 &  2.4  & 2157 \\
A1185-1 & 0395 & 23100 & 0.91 & 36.310 & 19.78 &  1.5  & 1423 \\
A1314-1 & 0395 & 19500 & 1.03 & 36.120 & 21.21 &  2.4  & 2222 \\
A1367-1 & 0295 &  4200 & 0.77 & 34.501 & 20.77 &  2.7  & 2113 \\
A1656-1 & 0593 &  4900 & 0.88 & 33.392 & 19.02 &  0.9  & 1216 \\
A1656-2 & 0593 &  5500 & 0.83 & 33.542 & 19.10 &  1.1  & 1663 \\
A1656-3 & 0395 & 10800 & 1.00 & 35.479 & 19.95 &  2.0  & 1429 \\
A2162-1 & 0995 & 18900 & 1.29 & 35.789 & 20.59 &  1.3  & 1196 \\
A2197-1 & 0395 & 19300 & 0.99 & 36.127 & 20.03 &  1.8  & 1210 \\
A2197-2 & 0395 & 12800 & 1.05 & 35.689 & 21.26 &  2.5  & 1786 \\
A2199-1 & 0794 & 11700 & 0.92 & 34.135 & 19.82 &  1.1  & 2530 \\
A2634-1 & 0995 & 13700 & 1.01 & 35.428 & 20.04 &  1.4  & 1455 \\
A2666-1 & 1194 &  8200 & 0.91 & 35.023 & 20.83 &  2.4  & 1036 \\
\enddata
\end{deluxetable}

\begin{deluxetable}{rcrrrrcrrrr}
\small \centering\tablewidth{0pt}
\tablecaption{Point Source Counts and Variance Measurements\label{tab:datab}}
\newdimen\digitwidth
\newdimen\minuswidth
\setbox0=\hbox{\rm0}
\digitwidth=\wd0
\catcode`?=\active
\def?{\kern\digitwidth}
\tablehead{
\colhead{Galaxy.reg} & \colhead{?$m_b$} & \colhead{$N_{ps}$} & \colhead{$\pm$}
& \colhead{$N_{GC}$} & \colhead{$\pm$} & \colhead{$m_c$} & \colhead{$P_0$} &
\colhead{$\pm$} & \colhead{$P_{GC}$} & \colhead{$\pm$} }
\startdata
\multicolumn{11}{c}{Runs 0794 \& 1194.}\\ 
\noalign{\vskip 2pt}\hline
 A262-1.c1 &  21.0&  97.6& 28.1&  76.6& 28.3& 24.0&  1090&   53&   931&   75 \nl
 A262-1.c2 &  21.0&  76.3& 10.1&  55.3& 10.8& 24.0&   546&   24&   415&   57 \nl
 A262-1.c3 &  21.0&  50.9&  4.6&  29.9&  5.9& 24.0&   371&   15&   253&   53 \nl
 A262-1.c4 &  21.0&  27.2&  1.8&   6.2&  4.1& 24.0&   201&   18&    88&   54 \nl
 A347-1.c1 &  21.0& 122.8& 26.3& 102.0& 26.4& 24.0&   688&   39&   606&   43 \nl
 A347-1.c2 &  21.0&  54.9&  8.4&  34.1&  8.6& 24.0&   313&   17&   243&   24 \nl
 A347-1.c3 &  21.0&  30.5&  3.2&   9.7&  3.7& 24.0&   189&    4&   123&   17 \nl
 A347-1.c4 &  21.0&  22.5&  1.4&   1.7&  2.3& 24.0&   103&    7&    39&   18 \nl
A2199-1.c2 &  21.5& 100.0& 15.6&  74.0& 15.7& 24.0&   311&   29&   262&   31 \nl
A2199-1.c3 &  21.5&  68.9&  6.0&  42.9&  6.2& 24.0&   151&    8&   107&   12 \nl
A2199-1.c4 &  21.5&  36.1&  2.1&  10.1&  2.6& 24.0&    89&    6&    48&   11 \nl
A2666-1.c1 &  21.5&  94.5& 38.8&  76.1& 38.9& 24.5&   848&  115&   712&  121 \nl
A2666-1.c2 &  21.5&  27.7&  6.8&   9.3&  7.4& 24.5&   412&   23&   297&   43 \nl
A2666-1.c3 &  21.5&  57.1&  5.1&  24.8&  7.6& 25.0&   128&   13&    59&   26 \nl
A2666-1.c4 &  21.5&  39.9&  2.4&   7.6&  6.1& 25.0&    94&    7&    28&   24 \nl
\hline\noalign{\vskip 2pt}
\multicolumn{11}{c}{Run 0295.}\\
\noalign{\vskip 2pt}\hline
 A539-1.c1 &  22.0&  80.8& 24.4&  63.1& 24.7& 24.5&  1813&  214&  1539&  233 \nl
 A539-1.c2 &  22.0&  44.7&  7.8&  27.0&  8.6& 24.5&   775&   43&   517&  101 \nl
 A539-1.c3 &  22.0&  35.0&  4.0&  17.3&  5.4& 24.5&   520&   13&   267&   93 \nl
 A539-1.c4 &  22.0&  22.3&  1.5&   4.6&  3.9& 24.5&   285&   15&    34&   93 \nl
 A539-2.c1 &  22.0& 101.8& 31.1&  80.2& 31.2& 24.5&  2217&   76&  1954&  118 \nl
 A539-2.c2 &  22.0&  92.5& 13.5&  70.9& 13.7& 24.5&  1055&   26&   802&   94 \nl
 A539-2.c3 &  22.0&  44.9&  4.6&  23.3&  5.2& 24.5&   562&   18&   315&   92 \nl
 A539-2.c4 &  22.0&  24.1&  1.7&   2.5&  3.0& 24.5&   326&   17&    82&   92 \nl
 A569-1.c1 &  21.5&  16.1&  9.4&   4.7&  9.6& 24.0&   336&   31&   260&   38 \nl
 A569-1.c2 &  21.5&  68.2& 11.0&  45.1& 11.2& 24.5&    94&    7&    53&   15 \nl
 A569-1.c3 &  21.5&  35.3&  3.6&  12.2&  4.3& 24.5&    49&    2&    12&   13 \nl
 A569-1.c4 &  21.5&  24.5&  1.6&   1.4&  2.8& 24.5&    44&    2&     8&   13 \nl
 A634-1.c1 &  22.0& 145.8& 32.4& 128.0& 32.4& 24.5&  2282&  118&  1807&  145 \nl
 A634-1.c2 &  22.0&  48.1&  9.1&  30.3&  9.2& 24.5&  1023&   62&   591&  104 \nl
 A634-1.c3 &  22.0&  50.3&  4.2&  13.7&  4.6& 25.0&   318&   10&    54&   53 \nl
 A634-1.c4 &  22.0&  38.4&  2.0&   1.8&  2.7& 25.0&   237&   13& $-$25&   54 \nl
 A779-1.c1 &  21.5&  86.6& 36.2&  68.1& 36.4& 24.5&  1083&   59&   940&   65 \nl
 A779-1.c2 &  21.5& 113.7& 14.8&  78.2& 15.6& 25.0&   309&   26&   227&   31 \nl
 A779-1.c3 &  21.5&  91.2&  6.3&  55.7&  7.9& 25.0&   188&    7&   118&   16 \nl
 A779-1.c4 &  21.5&  45.3&  2.2&   9.8&  5.3& 25.0&   112&    8&    47&   16 \nl
 A999-1.c1 &  22.0&  98.5& 36.6&  68.2& 36.6& 25.0&  1068&   46&   775&   81 \nl
 A999-1.c2 &  22.0&  60.1& 10.6&  29.8& 10.7& 25.0&   512&   31&   228&   74 \nl
 A999-1.c3 &  22.0&  35.7&  4.2&   5.4&  4.5& 25.0&   345&   25&    67&   71 \nl
 A999-1.c4 &  22.0&  54.6&  2.3&   3.9&  3.8& 25.5&   133&    8& $-$40&   42 \nl
A1016-1.c1 &  22.0& 119.2& 30.9&  92.5& 30.9& 25.0&   685&   29&   451&   58 \nl
A1016-1.c2 &  22.0&  88.4& 12.1&  42.6& 12.4& 25.5&   226&   11&    84&   33 \nl
A1016-1.c3 &  22.0&  53.1&  4.9&   7.3&  5.7& 25.5&   153&    5&    15&   31 \nl
A1016-1.c4 &  22.0&  49.8&  2.5&   4.0&  3.8& 25.5&   111&    4& $-$26&   31 \nl
A1177-1.c2 &  22.0&  34.1&  8.3&  18.5&  8.5& 24.5&  1796&   94&  1158&  150 \nl
A1177-1.c3 &  22.0&  86.4&  5.9&  30.9&  6.3& 25.5&   384&    7&   137&   46 \nl
A1177-1.c4 &  22.0&  58.3&  2.3&   2.8&  3.2& 25.5&   251&   13&     8&   47 \nl
A1367-1.c1 &  21.5& 201.3& 38.3& 185.6& 38.3& 24.5&   448&   32&   396&   35 \nl
A1367-1.c2 &  21.5& 103.5& 11.9&  87.8& 12.0& 24.5&   236&    8&   194&   15 \nl
A1367-1.c3 &  21.5&  76.3&  5.2&  43.3&  6.0& 25.0&    64&    4&    40&    9 \nl
A1367-1.c4 &  21.5&  47.9&  2.2&  14.9&  3.7& 25.0&    38&    1&    16&    8 \nl
\hline\noalign{\vskip 2pt}
\multicolumn{11}{c}{Runs 0395 \& 0995.}\\
\noalign{\vskip 2pt}\hline
 A347-1.c1 &  21.5& 226.0& 36.5& 199.3& 36.7& 24.5&   302&   21&   261&   22 \nl
 A347-1.c2 &  21.5& 104.3& 11.7&  77.6& 12.4& 24.5&   120&    7&    88&   10 \nl
 A347-1.c3 &  21.5&  62.5&  4.7&  35.8&  6.2& 24.5&    69&    1&    40&    7 \nl
 A347-1.c4 &  21.5&  36.8&  1.9&  10.1&  4.5& 24.5&    41&    2&    13&    7 \nl
 A397-1.c1 &  22.0& 100.1& 26.0&  77.5& 26.1& 25.0&  1549&   71&  1288&   87 \nl
 A397-1.c2 &  22.0&  61.3&  9.7&  38.7&  9.9& 25.0&   886&   74&   653&   88 \nl
 A397-1.c3 &  22.0&  29.2&  3.3&   6.6&  3.8& 25.0&   548&   23&   324&   53 \nl
 A397-1.c4 &  22.0&  24.0&  1.4&   1.4&  2.3& 25.0&   331&   16&   111&   50 \nl
 A569-1.c1 &  21.5&  26.4& 13.9&   9.5& 14.1& 24.0&   621&   97&   485&  105 \nl
 A569-1.c2 &  21.5&  35.4&  7.1&  18.5&  7.5& 24.0&   317&   22&   202&   44 \nl
 A569-1.c3 &  21.5&  21.6&  2.7&   4.7&  3.6& 24.0&   175&   20&    68&   42 \nl
 A569-1.c4 &  21.5&  17.6&  1.4&   0.7&  2.8& 24.0&   150&   16&    45&   41 \nl
A1185-1.c1 &  22.0& 126.8& 51.2&  99.0& 51.3& 25.0&  5627&  344&  4921&  367 \nl
A1185-1.c2 &  22.0&  62.1&  9.9&  34.3& 10.2& 25.0&  2327&  137&  1683&  184 \nl
A1185-1.c3 &  22.0&  39.6&  3.8&  11.8&  4.6& 25.0&  1377&   33&   755&  127 \nl
A1185-1.c4 &  22.0&  30.1&  1.9&   2.3&  3.2& 25.0&   889&   54&   274&  134 \nl
A1314-1.c1 &  22.0&  84.6& 28.8&  57.1& 28.9& 25.0&  2531&  239&  1898&  263 \nl
A1314-1.c2 &  22.0&  85.9& 12.3&  58.4& 12.6& 25.0&  1439&  123&   838&  164 \nl
A1314-1.c3 &  22.0&  51.6&  4.8&  24.1&  5.6& 25.0&   908&   34&   323&  113 \nl
A1314-1.c4 &  21.5&  64.6&  3.0&  20.4&  6.2& 25.5&   342&   12& $-$19&   68 \nl
A1656-3.c1 &  21.5& 152.5& 30.4& 127.5& 30.5& 24.5&  2417&  103&  2099&  136 \nl
A1656-3.c2 &  21.5&  62.9&  9.2&  37.9&  9.5& 24.5&  1323&   96&  1053&  128 \nl
A1656-3.c3 &  21.5&  36.1&  3.4&  11.1&  4.2& 24.5&   600&   22&   347&   88 \nl
A1656-3.c4 &  21.5&  27.5&  1.7&   2.5&  3.0& 24.5&   391&   21&   144&   88 \nl
A2197-1.c2 &  22.0&  39.3&  8.0&  21.0&  9.0& 24.5&  2511&  189&  1406&  430 \nl
A2197-1.c3 &  22.0&  25.3&  3.1&   7.0&  5.2& 24.5&  1318&  126&   252&  406 \nl
A2197-1.c4 &  22.0&  21.7&  1.3&   3.4&  4.4& 24.5&  1128&   58&    73&  390 \nl
A2197-2.c1 &  22.0& 161.8& 50.3& 138.3& 50.5& 25.0&  1451&   61&  1221&   79 \nl
A2197-2.c2 &  22.0&  81.7& 12.7&  58.2& 13.6& 25.0&   857&   56&   653&   74 \nl
A2197-2.c3 &  22.0&  60.0&  5.0&  36.5&  7.0& 25.0&   499&   15&   307&   51 \nl
A2197-2.c4 &  22.0&  32.0&  1.7&   8.5&  5.2& 25.0&   281&   14&    94&   51 \nl
A2162-1.c1 &  22.0&  55.7& 20.1&  45.1& 20.2& 24.0&  5886&  539&  5009&  569 \nl
A2162-1.c2 &  22.0&  71.3& 12.1&  50.9& 12.3& 24.5&  1775&  124&  1240&  168 \nl
A2162-1.c3 &  22.0&  37.4&  3.8&  17.0&  4.4& 24.5&   835&   40&   312&  120 \nl
A2162-1.c4 &  22.0&  23.7&  1.6&   3.3&  2.7& 24.5&   616&   34&    96&  118 \nl
A2634-1.c1 &  22.0& 110.3& 30.4&  91.5& 30.7& 24.5&  2435&  158&  2162&  173 \nl
A2634-1.c2 &  22.0&  75.6& 10.5&  56.8& 11.3& 24.5&  1207&   78&   960&  104 \nl
A2634-1.c3 &  22.0&  47.2&  4.2&  28.4&  5.9& 24.5&   738&   37&   501&   77 \nl
A2634-1.c4 &  22.0&  24.2&  1.6&   5.4&  4.4& 24.5&   429&   13&   195&   69 \nl
\enddata
\end{deluxetable}

\begin{deluxetable}{ccrc|ccrc}
\small \centering\tablewidth{0pt}
\tablecaption{Metric Specific Frequencies and GCLF Widths in CMB Frame\label{tab:snsig_cmb}}
\newdimen\digitwidth
\newdimen\minuswidth
\setbox0=\hbox{\rm0}
\digitwidth=\wd0
\catcode`?=\active
\def?{\kern\digitwidth}
\tabcolsep=0.3cm
\tablehead{
\colhead{Galaxy} & \colhead{$M_V$} & \colhead{$S_N\;\,^+_-\,$}  & \colhead{$\;\sigma_{\rm LF}\;\;^+_-\,$}  &
\colhead{Galaxy} & \colhead{$M_V$} & \colhead{$S_N\;\,^+_-\,$}  & \colhead{$\;\sigma_{\rm LF}\;\;^+_-\,$} }
\startdata
?A262-1 & $-$22.08& $5.0\;^{1.5}_{1.3}$ &  $1.38\;^{.14}_{.13}$ & A1314-1 &$-$22.20& $4.2\;^{1.1}_{1.0}$ &  $1.58\;^{.21}_{.20}$ \\
?A347-1 & $-$21.86& $5.8\;^{1.6}_{1.3}$ &  $1.36\;^{.10}_{.11}$ & A1367-1 &$-$22.29& $5.3\;^{1.4}_{1.1}$ &  $1.53\;^{.11}_{.12}$ \\
?A397-1 & $-$22.02& $4.7\;^{1.4}_{1.1}$ &  $1.43\;^{.11}_{.12}$ & A1656-1 &$-$22.75& $5.7\;^{1.3}_{1.3}$ &  $1.37\;^{.12}_{.13}$ \\
?A539-1 & $-$21.73& $9.1\;^{3.0}_{2.6}$ &  $1.42\;^{.13}_{.12}$ & A1656-2 &$-$22.55& $9.3\;^{2.0}_{2.0}$ &  $1.43\;^{.09}_{.09}$ \\
?A539-2 & $-$22.16& $9.4\;^{3.0}_{2.4}$ &  $1.46\;^{.12}_{.11}$ & A1656-3 &$-$22.16& $4.6\;^{1.5}_{1.3}$ &  $1.33\;^{.11}_{.11}$ \\
?A569-1 & $-$21.59& $3.0\;^{1.2}_{1.0}$ &  $1.42\;^{.24}_{.22}$ & A2162-1 &$-$21.98& $7.4\;^{2.2}_{1.8}$ &  $1.73\;^{.16}_{.15}$ \\
?A634-1 & $-$21.65& $4.0\;^{1.2}_{1.0}$ &  $1.76\;^{.18}_{.17}$ & A2197-1 &$-$22.16& $2.5\;^{1.4}_{1.3}$ &  $1.55\;^{.31}_{.24}$ \\
?A779-1 & $-$22.56& $4.1\;^{1.0}_{0.9}$ &  $1.34\;^{.12}_{.11}$ & A2197-2 &$-$22.22& $5.9\;^{1.5}_{1.2}$ &  $1.49\;^{.11}_{.10}$ \\
?A999-1 & $-$21.55& $3.9\;^{1.5}_{1.3}$ &  $1.56\;^{.24}_{.23}$ & A2199-1 &$-$22.44& $8.1\;^{2.3}_{1.9}$ &  $1.53\;^{.10}_{.09}$ \\
A1016-1 & $-$21.33& $3.3\;^{1.2}_{1.1}$ &  $1.83\;^{.31}_{.24}$ & A2634-1 &$-$22.40& $7.5\;^{2.1}_{1.7}$ &  $1.51\;^{.10}_{.09}$ \\
A1177-1 & $-$22.00& $4.2\;^{1.3}_{1.0}$ &  $1.49\;^{.18}_{.16}$ & A2666-1 &$-$22.24& $3.5\;^{1.1}_{1.0}$ &  $1.43\;^{.24}_{.19}$ \\
A1185-1 & $-$22.03& $6.4\;^{1.8}_{1.4}$ &  $1.33\;^{.10}_{.11}$ &         &        &              &                       \\
\enddata
\tablecomments{Columns list: galaxy name; absolute $V$ magnitude of the region of the
galaxy in which the GC measurements were made (assumes $H_0 = 80$ and CMB velocities);
\sn\ calculated within 40~kpc assuming 
$\sigma{=}1.40\pm0.05$ for the GCLF width; actual measured value of $\sigma$.}
\end{deluxetable}

\begin{deluxetable}{crcrcc|crcrcc}
\small \centering\tablewidth{0pt}
\tablecaption{Comparison of $S_N^{flu}$ and $S_N^{cnt}$ for $\sigma{\,=\,}$1.40~mag
\label{tab:snflucount}}
\newdimen\digitwidth
\newdimen\minuswidth
\setbox0=\hbox{\rm0}
\digitwidth=\wd0
\catcode`?=\active
\def?{\kern\digitwidth}
\tablehead{
\colhead{Galaxy} & \colhead{$S_N^{flu}$} &\colhead{$\pm$} & \colhead{$S_N^{cnt}$}
& \colhead{$\pm$} & \colhead{$\sigma_{dev}$} & \colhead{?Galaxy} & \colhead{$S_N^{flu}$}
& \colhead{$\pm$} & \colhead{$S_N^{cnt}$} & \colhead{$\pm$} & \colhead{$\sigma_{dev}$} }
\startdata
?A262-1 & 5.6 & 1.6 & 4.7 & 1.1  & 0.50 & A1185-1 &  6.7 & 0.7 & 5.4 & 1.3  & 0.96 \\
?A347-1 & 6.8 & 1.3 & 5.0 & 1.1  & 1.18 & A1314-1 &  3.4 & 0.7 & 6.7 & 1.1  & 2.65 \\
?A397-1 & 5.3 & 0.6 & 3.6 & 0.9  & 1.75 & A1367-1 &  5.6 & 1.0 & 6.0 & 0.6  & 0.35 \\
?A539-1 & 8.2 & 1.9 & 10.3 & 2.4 & 0.69 & A1656-3 &  5.7 & 1.1 & 3.8 & 1.0  & 1.29 \\
?A539-2 & 8.7 & 1.5 & 11.0 & 1.7 & 1.00 & A2162-1 &  6.3 & 1.1 & 15.5 & 2.8 & 3.17 \\
?A569-1 & 3.9 & 2.1 & 2.7 & 1.5  & 0.46 & A2197-1 &  1.8 & 1.5 & 3.7 & 1.9  & 0.80 \\
?A634-1 & 2.7 & 1.2 & 6.6 & 1.3  & 2.28 & A2197-2 &  5.5 & 0.7 & 7.1 & 1.1  & 1.22 \\
?A779-1 & 4.4 & 0.6 & 3.7 & 0.6  & 0.96 & A2199-1 &  7.5 & 0.8 & 11.2 & 1.4 & 2.37 \\
?A999-1 & 3.5 & 1.7 & 4.8 & 1.7  & 0.57 & A2634-1 &  7.1 & 0.8 & 9.4 & 1.5  & 1.43 \\
A1016-1 & 2.6 & 1.7 & 5.0 & 1.3  & 1.14 & A2666-1 &  3.7 & 1.0 & 3.0 & 0.9  & 0.53 \\
A1177-1 & 3.9 & 0.8 & 4.7 & 1.1  & 0.64 &         &      &     &     &      &     \\
\enddata
\end{deluxetable}

\begin{deluxetable}{ccrc|ccrc}
\small \centering\tablewidth{0pt}
\tablecaption{Metric Specific Frequencies and GCLF Widths in ACI Frame\label{tab:snsig_aci}}
\newdimen\digitwidth
\newdimen\minuswidth
\setbox0=\hbox{\rm0}
\digitwidth=\wd0
\catcode`?=\active
\def?{\kern\digitwidth}
\tabcolsep=0.3cm
\tablehead{
\colhead{Galaxy} & \colhead{$M_V$} & \colhead{$S_N\;\,^+_-\,$}  & \colhead{$\;\sigma_{\rm LF}\;\;^+_-\,$}  &
\colhead{Galaxy} & \colhead{$M_V$} & \colhead{$S_N\;\,^+_-\,$}  & \colhead{$\;\sigma_{\rm LF}\;\;^+_-\,$} }
\startdata
?A262-1 &$-$22.56& $ 5.4\;^{1.8}_{1.5}$ & $1.53\;^{.14}_{.13}$ & A1314-1 &$-$22.32& $ 4.3\;^{1.2}_{1.0}$  &  $1.62\;^{.21}_{.19}$ \\
?A347-1 &$-$22.28& $ 6.3\;^{1.8}_{1.5}$ & $1.48\;^{.10}_{.10}$ & A1367-1 &$-$22.28& $ 5.3\;^{1.4}_{1.1}$  &  $1.53\;^{.11}_{.12}$ \\
?A397-1 &$-$22.39& $ 5.2\;^{1.7}_{1.3}$ & $1.53\;^{.10}_{.11}$ & A1656-1 &$-$22.75& $ 5.7\;^{1.3}_{1.3}$  &  $1.37\;^{.12}_{.13}$ \\
?A539-1 &$-$22.07& $10.1\;^{3.5}_{2.9}$ & $1.51\;^{.13}_{.12}$ & A1656-2 &$-$22.55& $ 9.3\;^{2.0}_{2.0}$  &  $1.43\;^{.09}_{.09}$ \\
?A539-2 &$-$22.50& $10.5\;^{3.5}_{2.8}$ & $1.55\;^{.12}_{.12}$ & A1656-3 &$-$22.16& $ 4.6\;^{1.5}_{1.3}$  &  $1.33\;^{.11}_{.11}$ \\
?A569-1 &$-$21.87& $ 3.3\;^{1.2}_{1.0}$ & $1.50\;^{.24}_{.22}$ & A2162-1 &$-$22.01& $ 7.5\;^{2.3}_{1.9}$  &  $1.74\;^{.16}_{.16}$ \\
?A634-1 &$-$21.86& $ 4.1\;^{1.3}_{1.1}$ & $1.81\;^{.18}_{.16}$ & A2197-1 &$-$22.21& $ 2.6\;^{1.4}_{1.3}$  &  $1.56\;^{.30}_{.24}$ \\
?A779-1 &$-$22.70& $ 4.1\;^{1.0}_{0.9}$ & $1.39\;^{.12}_{.12}$ & A2197-2 &$-$22.28& $ 5.9\;^{1.5}_{1.2}$  &  $1.51\;^{.11}_{.11}$ \\
?A999-1 &$-$21.67& $ 4.0\;^{1.6}_{1.4}$ & $1.59\;^{.24}_{.23}$ & A2199-1 &$-$22.50& $ 8.2\;^{2.4}_{1.9}$  &  $1.55\;^{.09}_{.09}$ \\
A1016-1 &$-$21.45& $ 3.3\;^{1.2}_{1.1}$ & $1.86\;^{.31}_{.23}$ & A2634-1 &$-$22.72& $ 8.1\;^{2.4}_{1.9}$  &  $1.61\;^{.10}_{.10}$ \\
A1177-1 &$-$22.08& $ 4.3\;^{1.3}_{1.1}$ & $1.51\;^{.18}_{.16}$ & A2666-1 &$-$22.58& $ 3.8\;^{1.3}_{1.1}$  &  $1.53\;^{.23}_{.19}$ \\
A1185-1 &$-$22.11& $ 6.5\;^{1.9}_{1.5}$ & $1.35\;^{.10}_{.10}$ &         &        &                &                       \\
\enddata
\tablecomments{\small 
Same as Table~\ref{tab:snsig_cmb} but all calculations done in the ACI frame.}
\end{deluxetable}

\begin{table}
\caption{ Neighboring Galaxy Counts\label{tab:jbcounts}}
\vspace{3mm}
\small
\newdimen\digitwidth
\newdimen\minuswidth
\setbox0=\hbox{\rm0}
\digitwidth=\wd0
\catcode`?=\active
\def?{\kern\digitwidth}
\def\jj{\llap{$\ge$}}
\tabcolsep=0.3cm
\begin{tabular}{c|ccc|cc}
\hline\hline 
\multicolumn{1}{r|}{$L >$} &\multicolumn{3}{c}{?$.10\,L^*$} &\multicolumn{2}{|c}{?$.05\,L^*$}\\
\multicolumn{1}{r|}{$r <$}& ?32 & 40 & 50 & ?32 & 40 \rlap{\hbox{\phn(\hkpc)}}\\
Galaxy& $?N_n$ & $N_n$ & $N_n$ & ?$N_n$ & $N_n$ \\
\tableline
?A262-1 & ?3 & 5 &\jj5 &   ?5  &  7  \\
?A347-1 & ?0 & 0 &\jj0 &   ?0  &  0  \\
?A397-1 & ?2 & 2 & 2   &   ?4  &  4  \\
?A539-1 & ?2 & 4 & 4   &   ?2  &  4  \\
?A539-2 & ?4 & 6 & 7   &   ?6  &  8  \\
?A569-1 & ?0 & 1 & 3   &   ?0  &  1  \\
?A634-1 & ?0 & 1 & 1   &   ?0  &  1  \\
?A779-1 & ?2 & 3 & 3   &   ?3  &  4  \\
?A999-1 & ?2 & 4 & 4   &   ?3  &  6  \\
A1016-1 & ?1 & 2 & 2   &   ?1  &  2  \\
A1177-1 & ?0 & 2 & 2   &   ?1  &  4  \\
A1185-1 & ?3 & 3 & 3   &   ?4  &  5  \\
A1314-1 & ?0 & 1 & 3   &   ?1  &  2  \\
A1367-1 & ?3 & 3 & 3   &   ?4  &  5  \\
A1656-1 & ?2 & 4 & 6   &   ?3  &  5  \\
A1656-2 & ?5 & 7 &\llap{1}1  &   ?5  &  7  \\
A1656-3 & ?1 & 1 & 2   &   ?1  &  1  \\
A2162-1 & ?0 & 1 & 1   &   ?0  &  3  \\
A2197-1 & ?0 & 1 & 1   &   ?0  &  1  \\
A2197-2 & ?1 & 1 & 2   &   ?2  &  2  \\
A2199-1 & ?5 & 5 & 8   &   ?6  &  7  \\
A2634-1 & ?2 & 3 & 3   &   ?3  &  4  \\
A2666-1 & ?1 & 1 & 3   &   ?1  &  1  \\
\hline
\end{tabular}          
\end{table}

\clearpage

\clearpage

\clearpage

\end{document}